\documentclass{aastex62}

\graphicspath{{./}{figures/}}

\received{\today}
\revised{ }
\accepted{ }
\submitjournal{ApJ}

\shorttitle{Narrow RHESSI Flare}
\shortauthors{Dennis and Tolbert}
\begin{document}

\title{A  Very Narrow RHESSI X-ray Flare on 25 September 2011}

\correspondingauthor{Brian R. Dennis}
\email{brian.r.dennis@nasa.gov}

\author[0000-0001-8585-2349]{Brian. R. Dennis}
\affil{Solar Physics Laboratory, NASA Goddard Space Flight Center, Greenbelt, MD 20771, USA\\}

\author{Anne K. Tolbert}
\affiliation{Solar Physics Laboratory, NASA Goddard Space Flight Center, Greenbelt, MD 20771, USA\\}
 \affiliation{American University, Washington, DC}




\texttt{texcount -incbib -v3 NarrowFlare.tex})

\begin{abstract}

%The Astrophysical Journal (ApJ) has a 250 word limit for the abstract.  If you exceed this length the Editorial office will ask you to shorten it.

The unusually narrow X-ray source imaged with RHESSI during an impulsive spike lasting for $\sim$10~s during the GOES C7.9 flare on 25 September 2011 (SOL2011-09-25T03:32) was only $\sim$2~ arcsec wide and $\sim$10~arcsec long. Comparison with HMI magnetograms and AIA images at 1700~\AA~shows that the X-ray emission was primarily from a long ribbon in the region of positive polarity with little if any emission from the negative polarity ribbon. However, a thermal plasma source density of $\sim$10$^{12}~cm^{-3}$ estimated from the RHESSI-derived emission measure and source area showed that this could best be interpreted as a coronal hard X-ray source in which the accelerated electrons with energies lass than $\sim$50~keV were stopped by Coulomb collisions in the corona, thus explaining the lack of the more usual bright X-ray footpoints. Analysis of RHESSI spectra shows greater consistency with a multi-temperature distribution and a low energy cutoff to the accelerated electron spectrum of 22 keV compared to 12 keV if a single temperature distribution is assumed. This leads to a change in the lower limit on the total energy in electrons by an order of magnitude given the steepness of the best-fit electron spectrum with a power-law index of $\sim$6.  

%The RHESSI X-ray spectrum between 6 and 50 keV at the time of the impulsive hard X-ray peak lasting for $\sim$10 s at 03:30:18 UT is equally consistent with the sum of a nonthermal thick-target emission from electrons with a power-law spectrum with an index of -5.8$\pm$0.2 and three possible thermal functions - (1)~thermal emission from plasma with a single temperature of 0.9 keV (10 MK), (2)~two distinct temperatures, or (3)~a differential emission measure (DEM) that is a decreasing power-law function with increasing temperature with an index of $\sim$6.. Assuming that the impulsive peak seen at higher energies was nonthermal while the more gradually varying emission seen at lower energies was thermal, it was found that a multi-temperature model with a power-law DEM is more consistent with the observations than the other two assumptions. This multi-temperature model results in an increase by a factor of $\sim$2 in the upper limit determined for the low energy cutoff (E$_c$) to the spectrum of electrons that produce the nonthermal component - from a value of 12 keV assuming a single temperature plasma to 22 keV assuming a power-law DEM. This increase in E$_c$ results in a change in the lower limit on the total energy in electrons by an order of magnitude given the steepness of the best-fit electron spectrum.}

\end{abstract}



\section{Introduction} \label{sec:intro}

% Flares are not observed from inside sunspots despite the fact that they are regions with the strongest magnetic field but the lowest temperatures in the solar atmosphere. Even the small inverted Y-shaped structures suggestive of magnetic reconnection between magnetic dipoles and the unipolar background fields have only been identified in the penumbral region \citep{2016ApJ...819L...3Z} or in a penumbral intrusion into a sunspot umbra \citep{2017A&A...597A.127B}. Some sunspots have bright lanes across the dark umbra, called light bridges, and surge-like chromospheric activity has been observed above some light bridges. \cite{2018ApJ...854...92T} has reported the detection of ``frequently occurring magnetic reconnection and the resultant significant heating in the lower atmosphere of sunspot light bridges.''

The C7.9 flare of interest, designated as SOL2011-09-25T03:32, was first studied by \cite{Guo2012a,Guo2012,Guo2013}, who treated it as a coronal hard X-ray source since no obvious footpoints were detected. They fitted the observations with a collisional model with an extended electron acceleration region and a sufficiently high coronal density such that the accelerated electrons would lose all of their energy in the legs of the magnetic loop before reaching the chromospheric footpoints. The measured increase in source length with increasing photon energy was taken as evidence to quantify the parameters of this model.  However, \cite{2018ApJ...867...82D} pointed out that this interpretation was probably incorrect since the source was so narrow with a full width at half maximum (FWHM) of $^<_\sim$2~arcsec), and there was evidence of weak emission from at least one footpoint at high energies that would affect the measured energy dependence of the source length. Also, it was found that the linear X-ray source was near co-spatial with one of two possible ribbons apparent in the AIA 1700~\AA~images that themselves lay in a location with strong positive and negative magnetic fields situated between the two major sunspots of the active region.

This paper is a report on the detailed analysis of this unusual event with an emphasis on determining the physical nature of the X-ray source and the significance of its location with respect to the strong magnetic field magnitudes and gradients.  A comparison is presented in the appendix of nine image reconstruction algorithms available in the RHESSI software and of their ability to reproduce the finest features of this unusually narrow source.

\section{Observations}

\begin{figure}
    \centering 
    \includegraphics*[width=0.5\textwidth, angle = 90, 
   trim = 14 0 0 0]
   		{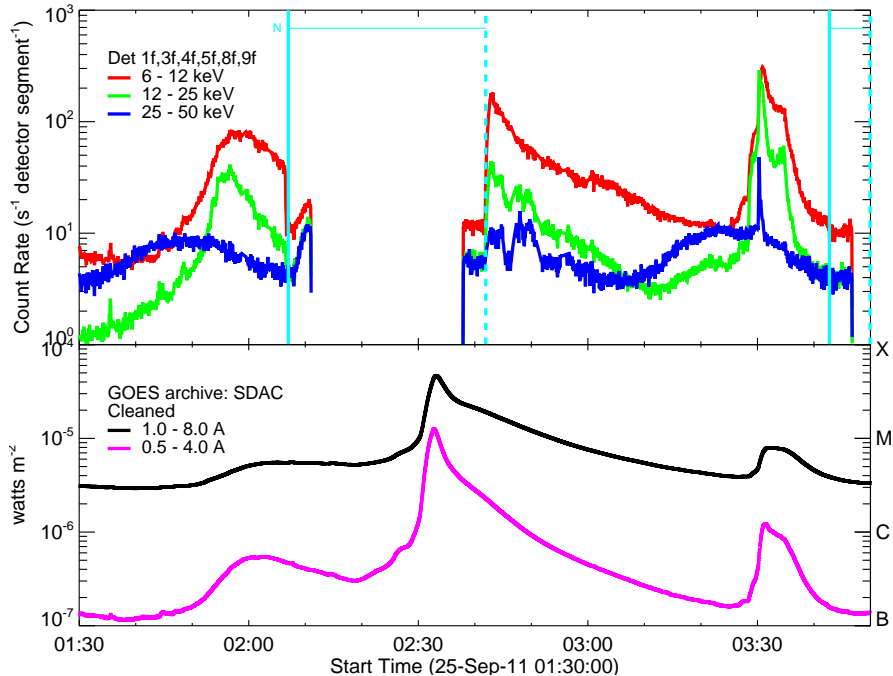}
    \caption{RHESSI and GOES lightcurves for two hours and twenty minutes showing the C7.9 flare peaking at 03:32 UT and the preceding two flares from different locations. The color-coded RHESSI curves are for the three indicated energy ranges. Summed count rates (corrected for attenuator changes) from the front segments of detectors 1, 3, 4, 5, 8, and 9 are plotted with a 4~s cadence to match the spacecraft spin period. The gradual peaks in the 25--50~keV lightcurve centered at 01:48 and 02:52 UT are not of solar origin but are caused by varying background rates as the spacecraft moves to higher geomagnetic latitudes. The data gaps between 02:07 and 02:42 UT and after 03:43 UT are the result of RHESSI night indicated by the cyan vertical and horizontal lines. The GOES two-channel lightcurves are shown in the lower plot for the same time interval with 2-s cadence.}
   \label{Fig-lc25sep2011}
   \end{figure}

The flare SOL2011-09-25T03:32 occurred on 25 September 2011 in NOAA active region 11302 at N12E50 (X=-700", Y=151"), peaking at 03:32~UT. The X-ray light curves from the Ramaty High Energy Solar Spectroscopic Imager (RHESSI) \citep{2002SoPh..210....3L} in three energy bins and from GOES are shown for a two hour period in Figure~\ref{Fig-lc25sep2011}. The C7.9 flare of interest is near the end of this time interval and appears as the impulsive spike in the 25 - 50 keV light curve matched by the rapid rise in the GOES fluxes at the same time. Two earlier peaks apparent in this figure are not relevant to the current discussion except that their decaying flux adds to the RHESSI count rates below $\sim$12 keV during SOL2011-09-25T03:32.  The earlier emission peaking at 01:57 UT was from AR11295 at X=884", Y=419". The GOES M4.4 event peaking at 02:33 UT was actually from two different active regions, one from the same active region as SOL2011-09-25T03:32 but the other from AR 11303 located at X=775", Y=-491".

\begin{figure}
    \centering
    \includegraphics[width=0.5\textwidth, angle = 90, 
   trim = 0 0 0 0]
   		{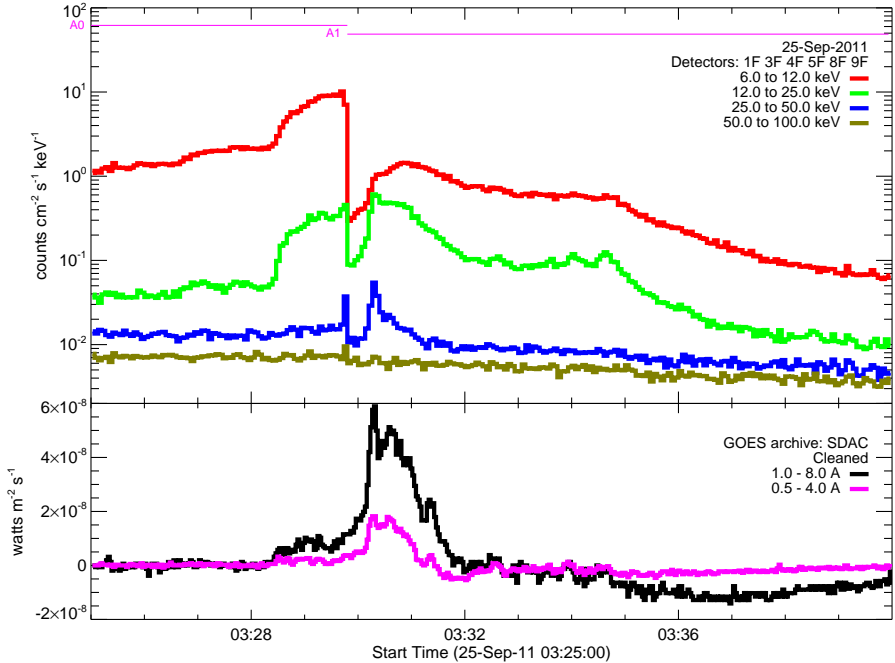}
    \caption{Zoomed in version of Figure \ref{Fig-lc25sep2011} showing RHESSI light curves in four energy ranges and GOES time-derivative light curves for the C7.9 flare. The RHESSI summed count fluxes with a 4-s cadence are corrected for live time but not for the insertion of the thin attenuators at 03:29:48 UT. The change from the A0 to the A1 attenuator state is shown by the solid pink lines at the top of the plot and by the drop in count rates at this time. The time derivatives of the two GOES lightcurves with 2-s cadence are shown in the lower plot.}
   \label{Fig-lc-RHESSI-GOESderiv}
   \end{figure}

More detailed RHESSI lightcurves of SOL2011-09-25T03:32 are shown in Figure~\ref{Fig-lc-RHESSI-GOESderiv}. The short impulsive peak at 03:31:18~UT can be seen most clearly in the 25--50~keV light curve. The decrease in the count rates at 03:29:48 UT is the result of the thin attenuators being inserted in the optical path of each detector at that time. Also shown are the time derivatives of the GOES light curves with an impulsive peak coincident with the impulsive HXR peak as expected from the Neupert Effect \citep{1968ApJ...153L..59N}.

\subsection{Imaging}

\begin{figure}
	\centering
    \includegraphics*[width=0.8\textwidth, angle = 90, 
   trim = 45 0 0 0]
   		{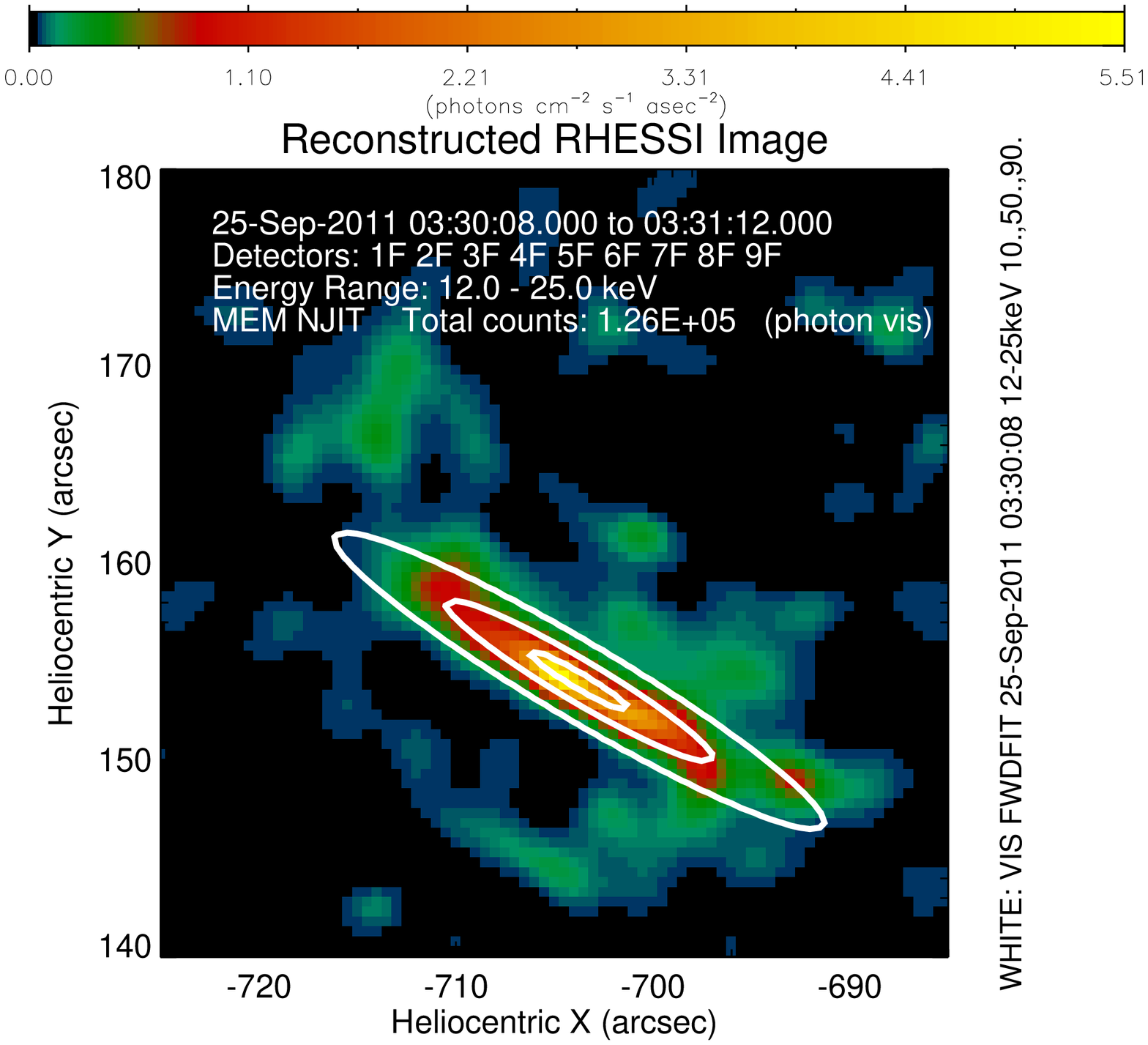}
    \caption{RHESSI color image in the 12--25~keV energy range made with the MEM\_NJIT reconstruction method for the 64~s duration of the 25--50~keV peak from 03:30:08 to 03:31:12 UT seen in Figure~\ref{Fig-lc-RHESSI-GOESderiv}.  The count rates of the front segments of all nine germanium detectors were used to make this image. The white contours are at 10, 50, and 90\% of the peak flux for the elliptical source reconstructed with the VIS\_FWDFIT algorithm. Note the possible evidence for separate emission seen as the compact red source at the extreme southwest end of the line feature at $X=-693,'' ~Y=149,''$ and the extension to the north in green at $X=-713,''~Y=168''$ that is also seen more convincingly at 6--12 keV (not shown).}
   \label{Fig-image1}
   \end{figure}

RHESSI imaging is based on the measurement of the solar X-ray flux that is modulated by nine bi-grid collimators as the spacecraft rotates \citep{2002SoPh..210...61H}. Various algorithms are available in the RHESSI software to reconstruct images from these measurements using different techniques (see Appendix in Section~\ref{Sec-Appendix} for a discussion of all currently available image reconstruction methods). Some algorithms, such as CLEAN and Pixon, use the modulated count rates in the different detectors directly, while others, such as MEM\_NJIT and VIS\_FWDFIT, use visibilities computed from the modulated  count rates in each detector. A visibility is defined as a vector representation of the amplitude and phase of the modulation at a given spacecraft roll angle with most instrumental artifacts removed \citep{2002SoPh..210...61H}.

{The RHESSI 12 --25~keV color image in Figure \ref{Fig-image1} reveals a long but remarkably narrow source with a width of $\sim$2~arcsec. The white contours show the corresponding image made with the VIS\_FWDFIT method under the assumption that the source is a single elliptical Gaussian. The best fit to the visibilities was obtained with a FWHM width of 2.0$\pm$0.2 arcsec and length of 15.8$\pm$0.4 arcsec.} Note that unless the assumption of two or more sources were made - an ellipse plus a circular Gaussian say - VIS\_FWDFIT cannot reveal the possible additional compact source suggested in the MEM\_NJIT image at the extreme southwest end (the compact source at X=-693,'' Y=149'') or the possible extension to the north. A double source assumption, however, did not lead to a significant reduction in the reduced $\chi^2$ value from that obtained with the single-ellipse assumption. Thus, the existence of the separate footpoint source suggested in the MEM\_NJIT image cannot be considered as statistically significant.

\begin{figure}
\centering
%	\plottwo{profile-nj-12-25kev}{profile-vf-12-25kev}
    \includegraphics[width=0.8\textwidth, angle = 0, 
    trim = 0 0 0 0]
    		{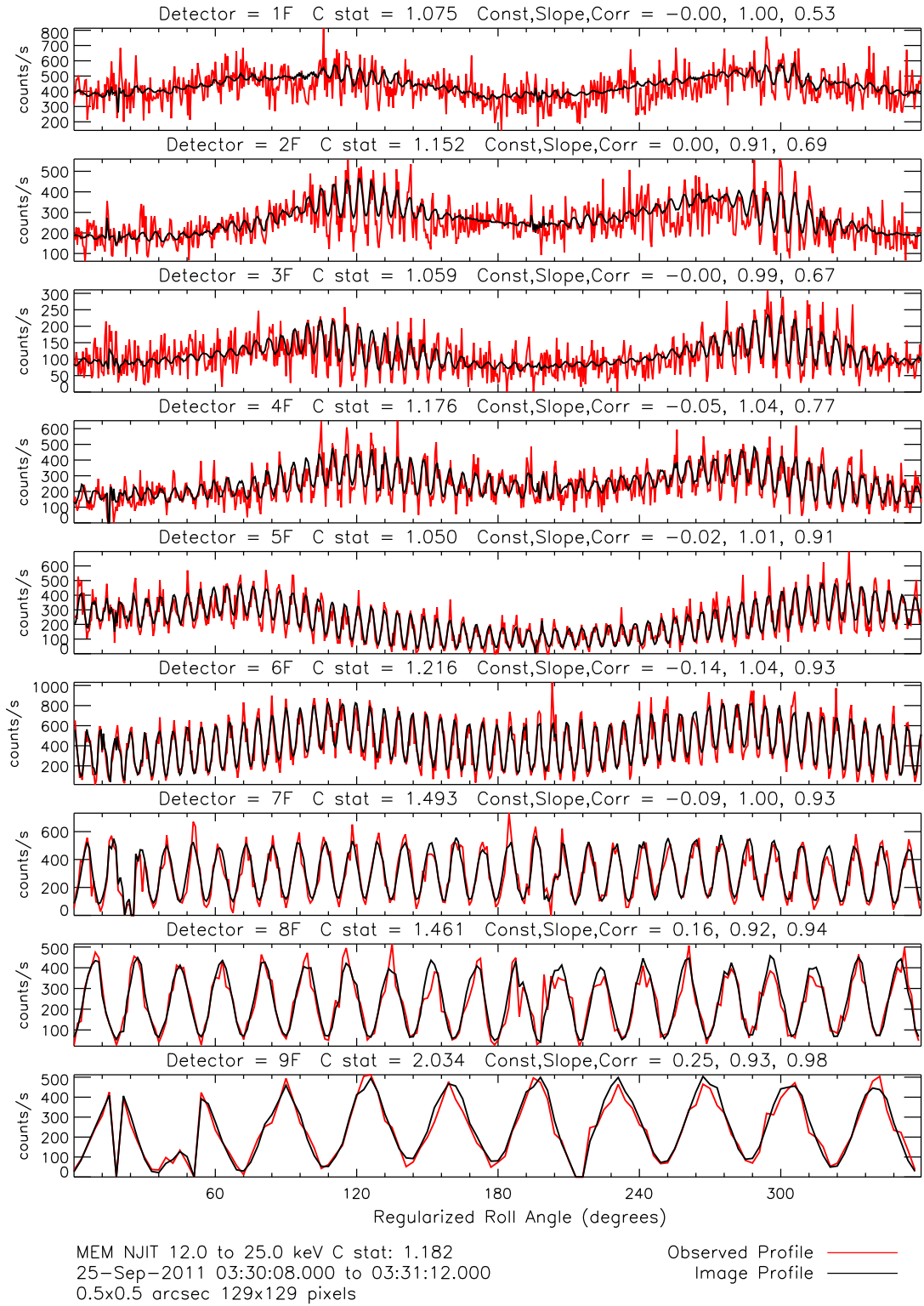}
    \caption{Measured count rates in red for the 12--25 keV energy band in the front segment of each of the nine germanium detectors plotted versus ``regularized roll angle.'' Overlaid in black are the count rates predicted from the image shown in Figure~\ref{Fig-image1} made using the MEM NJIT reconstruction algorithm. Both the measured and calculated count rates are accumulated modulo the spacecraft spin period for the duration of the 64~s time interval starting at 03:30:08~UT that was used to make the image. 
    }
   \label{Fig-profiles}
   \end{figure}

The existence of this long narrow source is revealed in the RHESSI light curves for the individual detectors at this time.  They show remarkably strong modulation in all detectors, including Detector \#1 that is behind a subcollimator with a nominal FWHM resolution of 2.3 arcsec. Figure \ref{Fig-profiles} shows the measured count rates (in red) in each detector and the predicted count rates (in black) from the reconstructed MEM NJIT image as a function of the ``regularized roll angle'' defined as the spacecraft roll angle corrected for the offset between the spin axis and the mean subcollimator optical axis.  For this plot, the count rates are summed modulo the spacecraft spin period ($\sim$4~s) for the 64~s interval duration used for the image. Significant sinusoidal modulation is evident in these plots for all detectors. The detectors behind the subcollimators with the coarser grids (\#5 to 9) show modulation at all roll angles while the detectors behind the finer grids (\#1 to 4) show modulation over two limited ranges in regularized roll angle ($\sim90^\circ$ to $150^\circ$ and $\sim280^\circ$ to $300^\circ$). This is the expected modulation signal for an asymmetrical source with a width much smaller than its length. The agreement between the measured and predicted rates is an indication of how well the reconstructed image matches the data. This agreement is quantified with the overall Cash or C-statistic\footnote{https://hesperia.gsfc.nasa.gov/$\sim$schmahl/cash/cash\_oddities.html} \citep{1979ApJ...228..939C} (as well as separate values for each detector) for comparison with values obtained with other reconstruction methods (see Appendix Section \ref{Sec-Appendix}).

\begin{figure}
    \includegraphics*[width=0.7\textwidth, angle = 90, 
   trim = 0 0 0 0]
   		{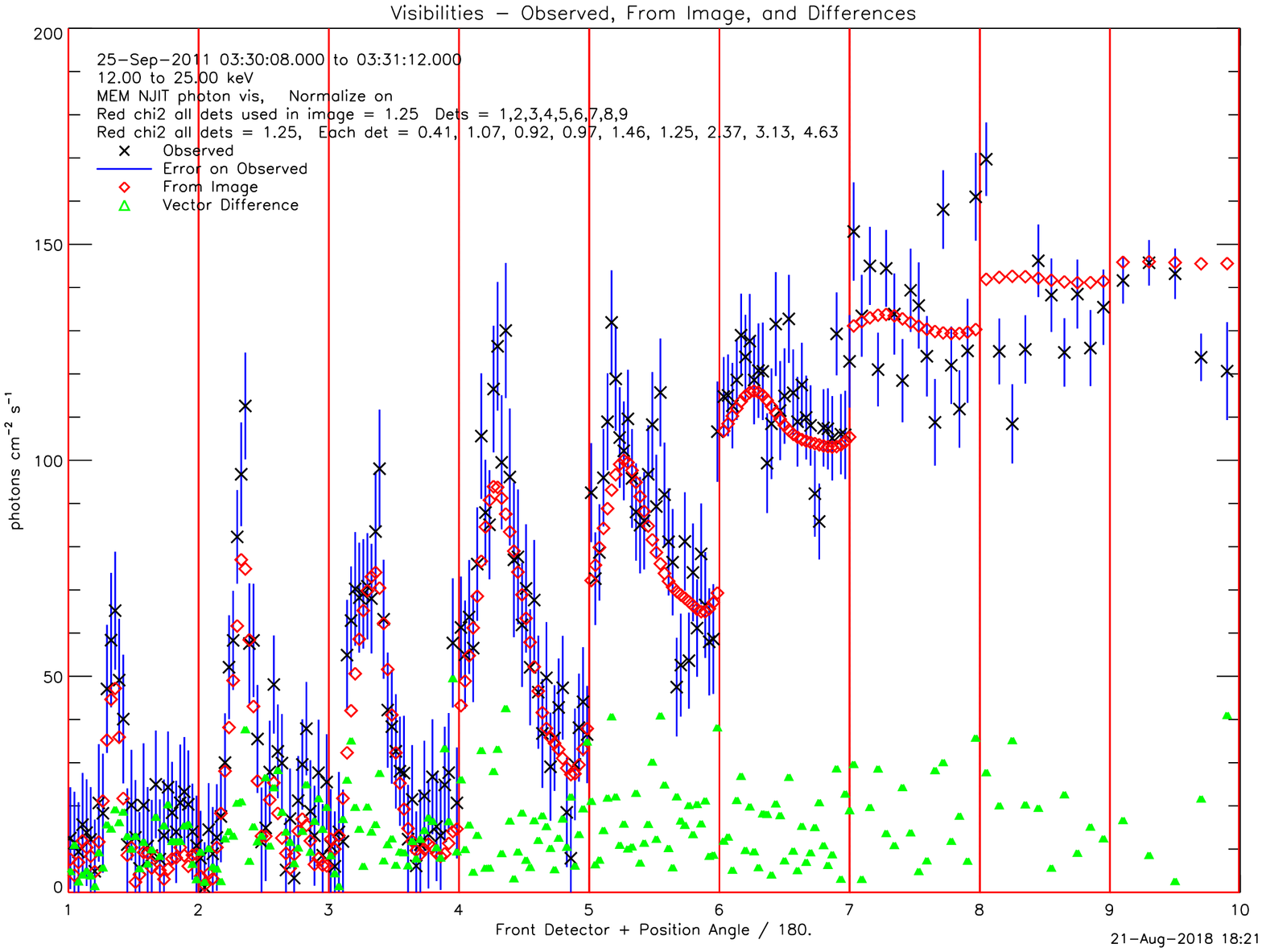}
    \caption{Visibility amplitudes plotted against position angle for each of the nine RHESSI detectors showing the modulation. The visibility amplitudes determined from the measured count rates are shown as black crosses with $\pm 1\sigma$~error bars in blue while the amplitudes determined from the image reconstructed using MEM\_NJIT are shown as red diamonds with the vector differences shown as green triangles.  The value of the reduced $\chi^2$ computed from the measured and calculated vector visibilities is 1.25 showing the remarkably good agreement. A similar plot was obtained for the VIS\_FWDFIT method. Note that the visibility amplitudes derived from the measured count rates have been converted to the plotted units of photons $cm^{-2} s^{-1}$ by multiplying by the diagonal elements of the individual detector response matrices relating the measured energy loss in the germanium detectors to the energy of an incident photon. }
   \label{Fig-visibilities}
   \end{figure}

Another way of showing the modulation in the different detectors is to plot the amplitude of the visibility vectors \citep{2002SoPh..210..165S,2007SoPh..240..241S} for each detector as a function of the position angle defined as the spatial direction of each grid response referenced to solar north. This is essentially the same as plotting just the amplitudes of the oscillations seen in Figure \ref{Fig-profiles} but here plotted for only a half rotation from 0 to 180$^\circ$ with the second half rotation assumed to be identical and added to the first, an option known as ``combine conjugates.''  Such a plot is shown in Figure \ref{Fig-visibilities}. Here again, the characteristics of an asymmetric source are dramatically evident with the peaks in the visibility amplitudes for each detector showing the position angle of the smallest dimension of the source and the valleys showing the position angle of the largest dimension. From this analysis, another quantitative indication of how well the image fits the count rates is given by the reduced $\chi^2$ values computed from the measured and predicted visibility vectors weighted by the statistical uncertainties. These values are listed in the figure for each detector and for all detectors together. The summed reduced $\chi^2$ values are used for comparison with values obtained for other reconstruction methods  (see Appendix Section~\ref{Sec-Appendix}).

The VIS\_FWDFIT method is unique amongst all the currently available reconstruction algorithms in giving source dimensions with uncertainties. The source flux, centroid location in X and Y, width, and length are determined that give the best fit to the measured visibilities using a weighted chi-squared value. The uncertainties are determined from a Monte Carlo analysis using 20 randomly chosen sets of visibilities distributed according to their statistical uncertainties. Images were made with this method in three energy ranges spanning the expected thermal and nonthermal domain between 6 and 50 keV with results listed in Table \ref{Tab-LoopWidth-Energy}. Surprisingly, neither the shape nor the dimensions of the source change appreciably with energy as the emission switches from predominantly thermal to nonthermal presumably at energies above $\sim$10 keV (see Section \ref{Spectroscopy}). The width perhaps decreases from 2.6 arcsec at low energies to below 2 arcsec in the 25--50 keV energy bin but no obvious footpoints show up and the source length stays constant at $\sim$15~arcsec. 

\begin{table}
\centering                  
\begin{tabular}{||c|c|c|c||}                                                                                                     
				\hline
				\hline   
		Energy& Width & Length  & Reduced\\
			(keV) & (arcsec) & (arcsec) &  $\chi^2$  \\
	\hline\hline
		6-12    & 2.6 $\pm$ 0.2  & 14.8 $\pm$ 0.4 & 1.6  \\ 
        		\hline
        12-25   & 2.0 $\pm$ 0.2  & 15.8 $\pm$ 0.4 & 1.4  \\  
				\hline                                                                                                                                   
		25-50   & 1.0 $\pm$ 0.2  & 15.4 $\pm$ 2.9 & 1.0  \\       
				\hline\hline  
\end{tabular}   
\caption{FWHM loop width and length with $\pm 1\sigma$ uncertainties for three different energy bins determined using the VIS\_FWDFIT image reconstruction algorithm with the assumption that the source was a single curved loop. The time interval used was from 03:30:08 to 03:31:12 UT to include the impulsive emission in the 25--50 keV light curve shown in Figures \ref{Fig-lc25sep2011} and~\ref{Fig-lc-RHESSI-GOESderiv}.
    }
\label{Tab-LoopWidth-Energy}
\end{table}

\begin{figure}
    \includegraphics*[width=0.45\textwidth, angle = 90, 
   trim = 20 60 00 50]
   		{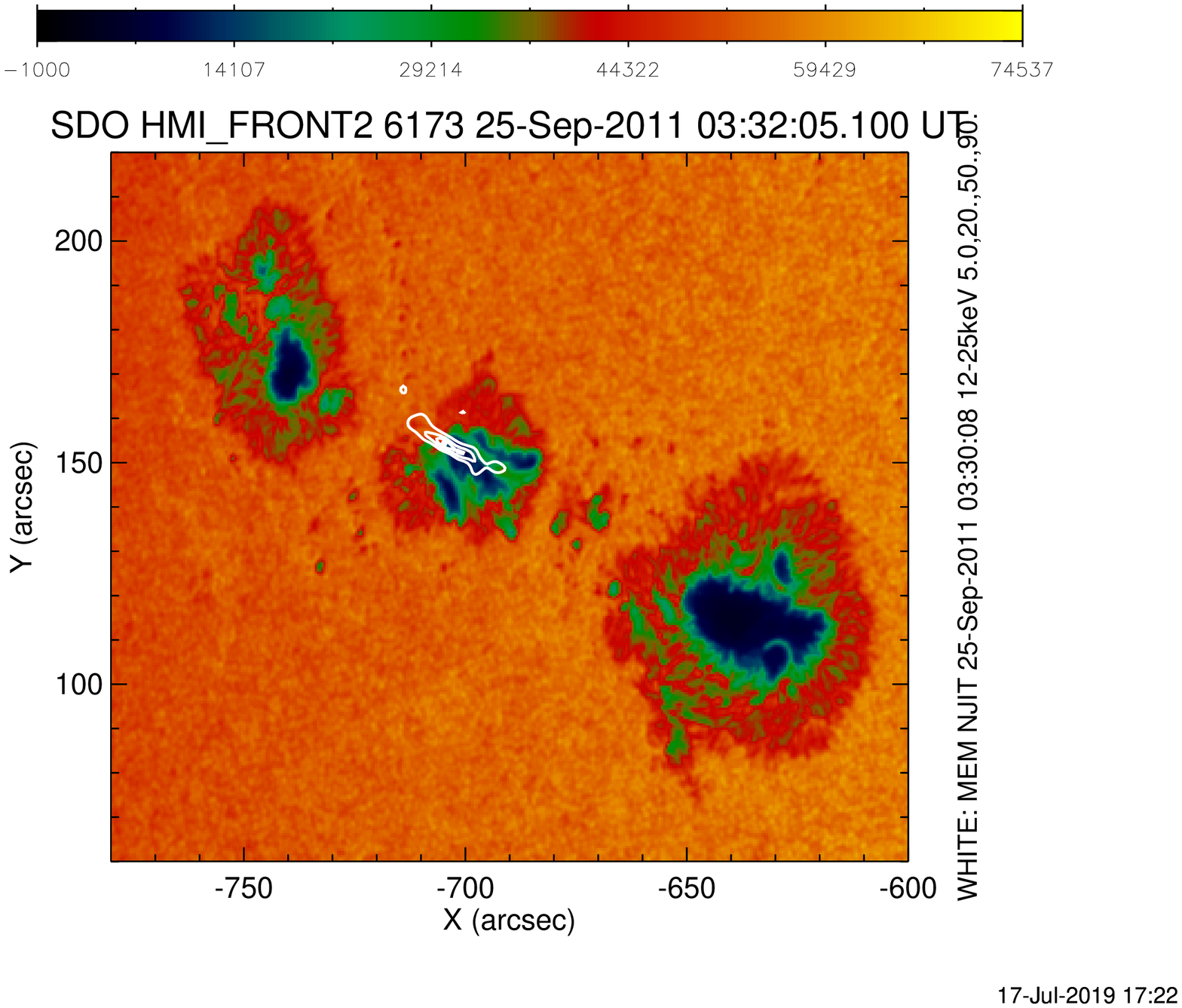}
   	 \includegraphics*[width=0.45\textwidth, angle = 90, 
   trim = 20 00 00 60]
   		{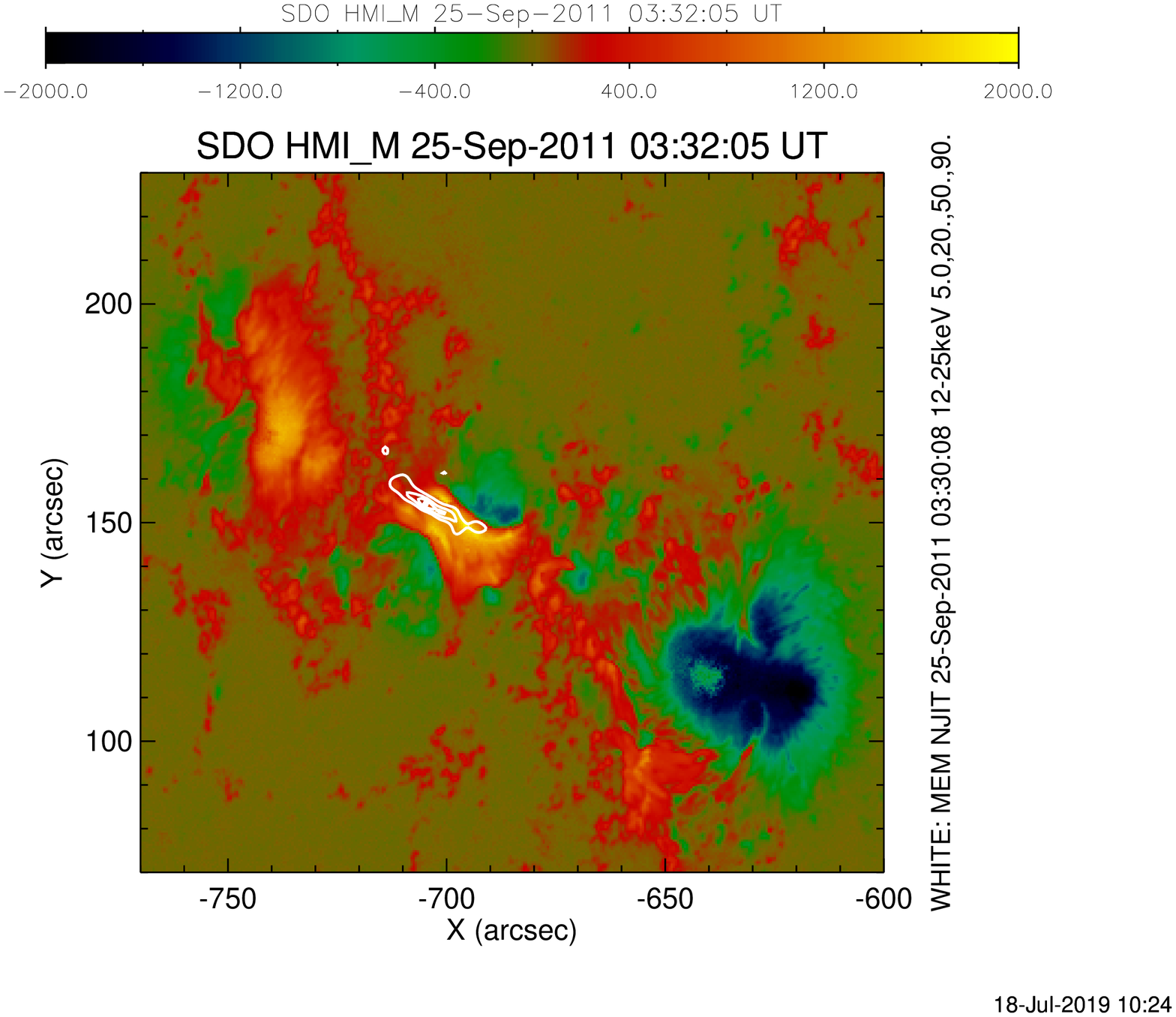}
    \caption{Left: HMI white-light image overlaid with RHESSI 12 - 25 keV flux contours at 5, 20, 50, and 90\% of the peak flux. The two main sunspots can be seen on either side of the central region where the flare occurred. Right: Same as the left image but using the HMI magnetogram instead of the white light image. }
   \label{Fig-im_HMI_RHESSI}
   \end{figure}

\begin{figure}
    \includegraphics*[width=0.45\textwidth, angle = 90, 
   trim = 20 80 0 60]
   		{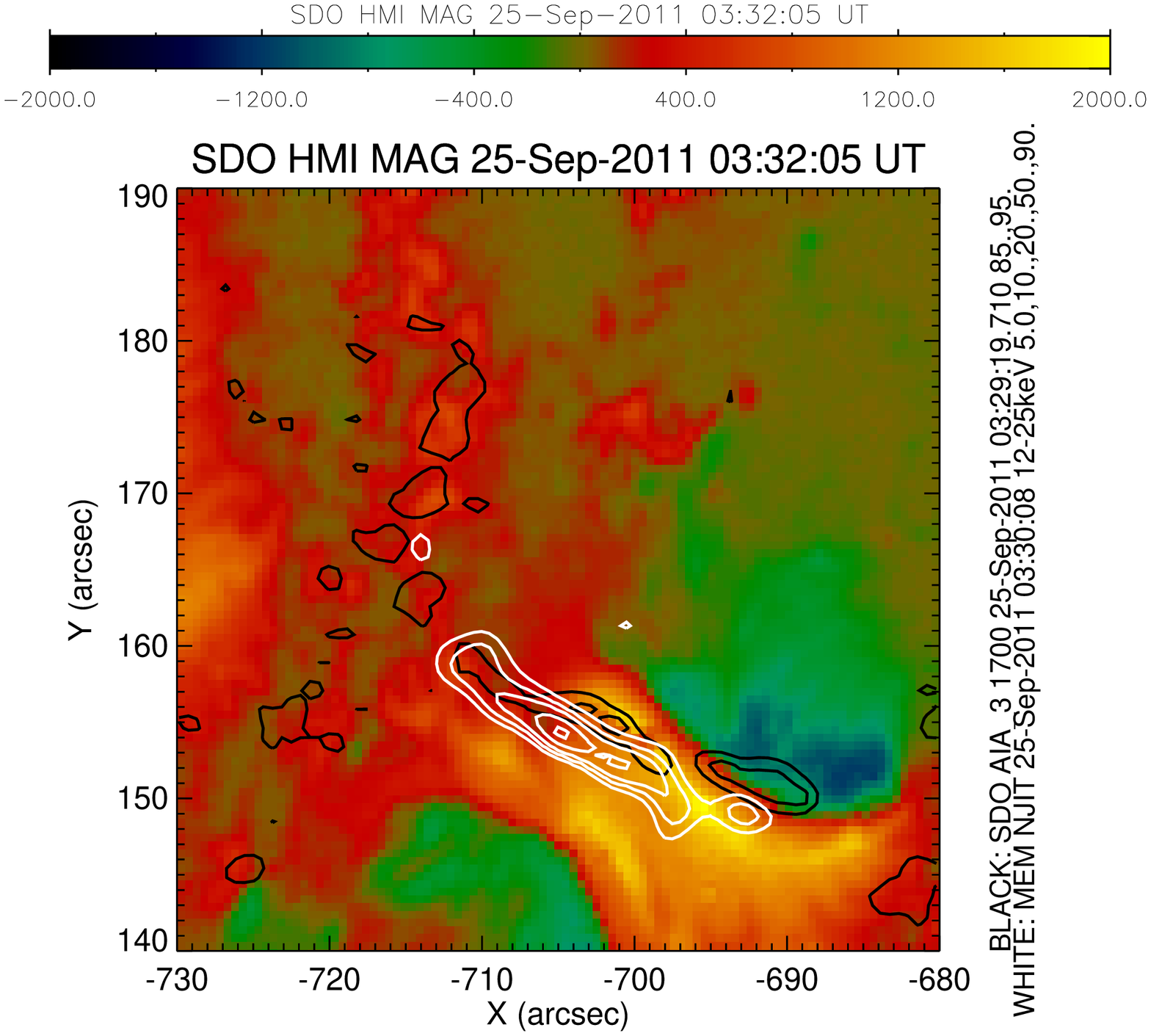}
   	\includegraphics*[width=0.45\textwidth, angle = 90, 
   trim = 20 0 0 60]
   		{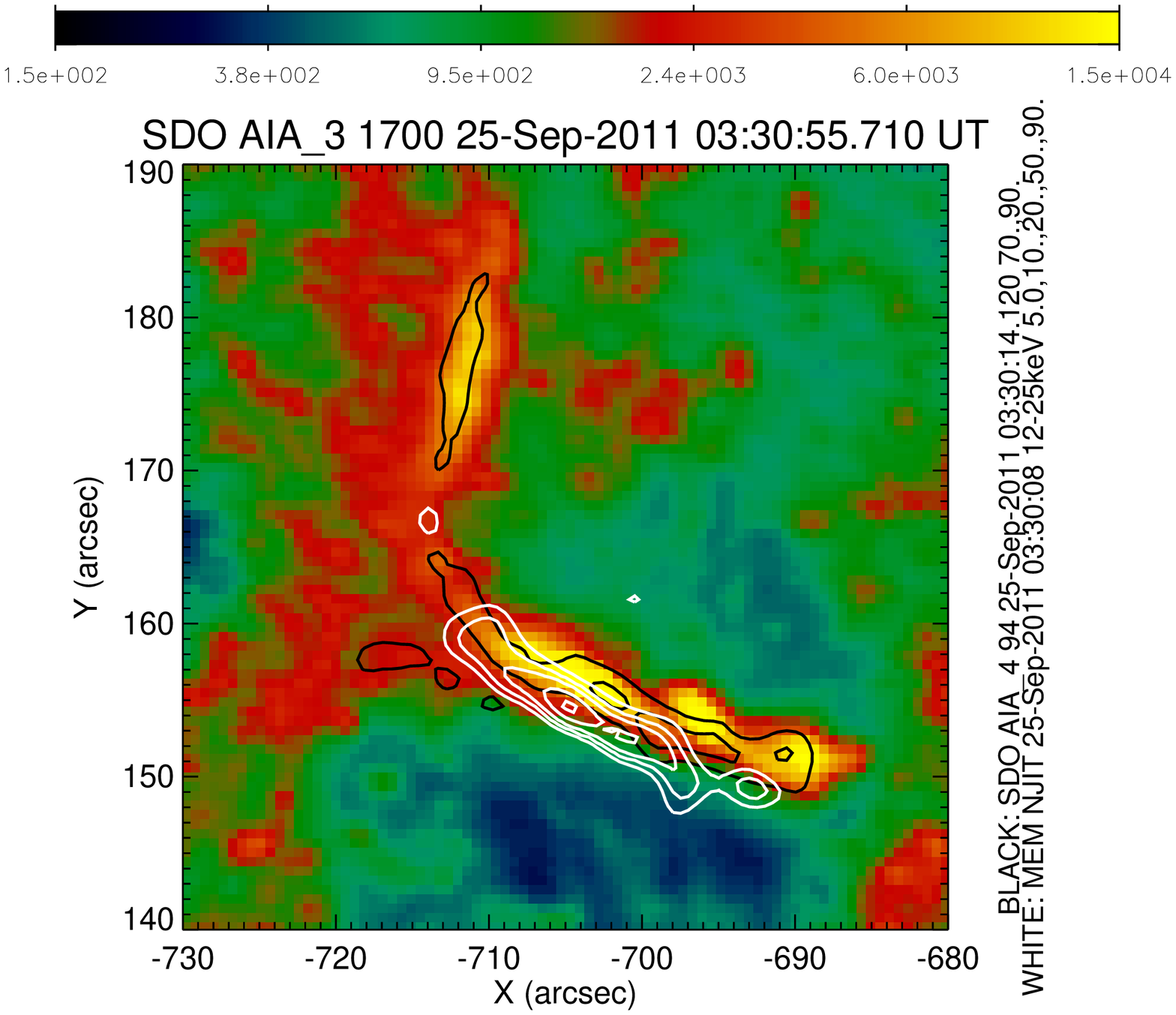}
    \caption{Left: A zoomed in version of the right panel of Figure~\ref{Fig-im_HMI_RHESSI} with AIA 1700~\AA~contours (black) added at 85, and 95\% of the peak flux. The central region can be seen with regions of strong, 2000 Gauss, positive and negative magnetic fields. with the neutral line bisecting it. The AIA 1700~\AA~contours show a double source separated by the neutral line with the eastern source extending to the north. The RHESSI 12--25~keV contours (white, 5, 10, 20, 50, and 90\%) also show a similar double source with a bright and very thin source in the region of positive polarity also extending up to the northeast and a much weaker compact source to the west. The  $\sim$2~arcsec displacement between the RHESSI sources and the 1700~\AA~sources suggests a roll error of 0.18$^\circ$ in either the RHESSI or the SDO aspect solutions. Right: Similar to the left panel showing an AIA 1700~\AA~image in color overlaid with AIA~94~\AA~70 and 90\%~contours (black) and the same RHESSI 12--25~keV contours (white).}
   \label{Fig-im_HMIMAG_AIA1700_RHESSI}
   \end{figure}

Overlaying RHESSI images of this flare on SDO/HMI magnetograms shows that it was located centrally between the two major sunspots in AR~11302 along the neutral line between oppositely directed magnetic fields. This is shown in Figure \ref{Fig-im_HMI_RHESSI}, where the RHESSI white contours show the location of the 12--25 keV emission. An expanded version is shown in Figure \ref{Fig-im_HMIMAG_AIA1700_RHESSI} with the addition of the SDO/AIA 1700~\AA~contours. Note that the AIA contours show two sources, one in the region of strong negative magnetic field and one in the region of positive magnetic field. This suggests that they are the two ribbons of the flare.  The RHESSI source lies mainly along the longer ribbon-like feature. Interestingly, the weak source to the southwest of the main line source in the MEM\_NJIT image shown in Figures \ref{Fig-image1} and \ref{Fig-im_HMIMAG_AIA1700_RHESSI} is located within $\sim$2 arcsec of the more compact source in the 1700~\AA~image. This is within the usual discrepancy found between RHESSI and AIA images.  

% \begin{figure}
%     \includegraphics*[width=0.3\textwidth, angle = 90, 
%   trim = 20 0 0 0]
%   		{image-HMI-mag_RHESSI.eps}
%     \caption{HMI magnetogram overlaid with RHESSI 12 - 25 keV flux contours at 5, 20, 50, and 90\% of the peak flux. The two main sunspots can be seen on either side of the central spot with the light bridge passing through it.  }
%   \label{Fig-im_HMI_RHESSI}
%   \end{figure}
   
\begin{figure}
    \centering
    \includegraphics*[width=0.5\textwidth, angle = 90, 
   trim = 0 0 25 0]
   		{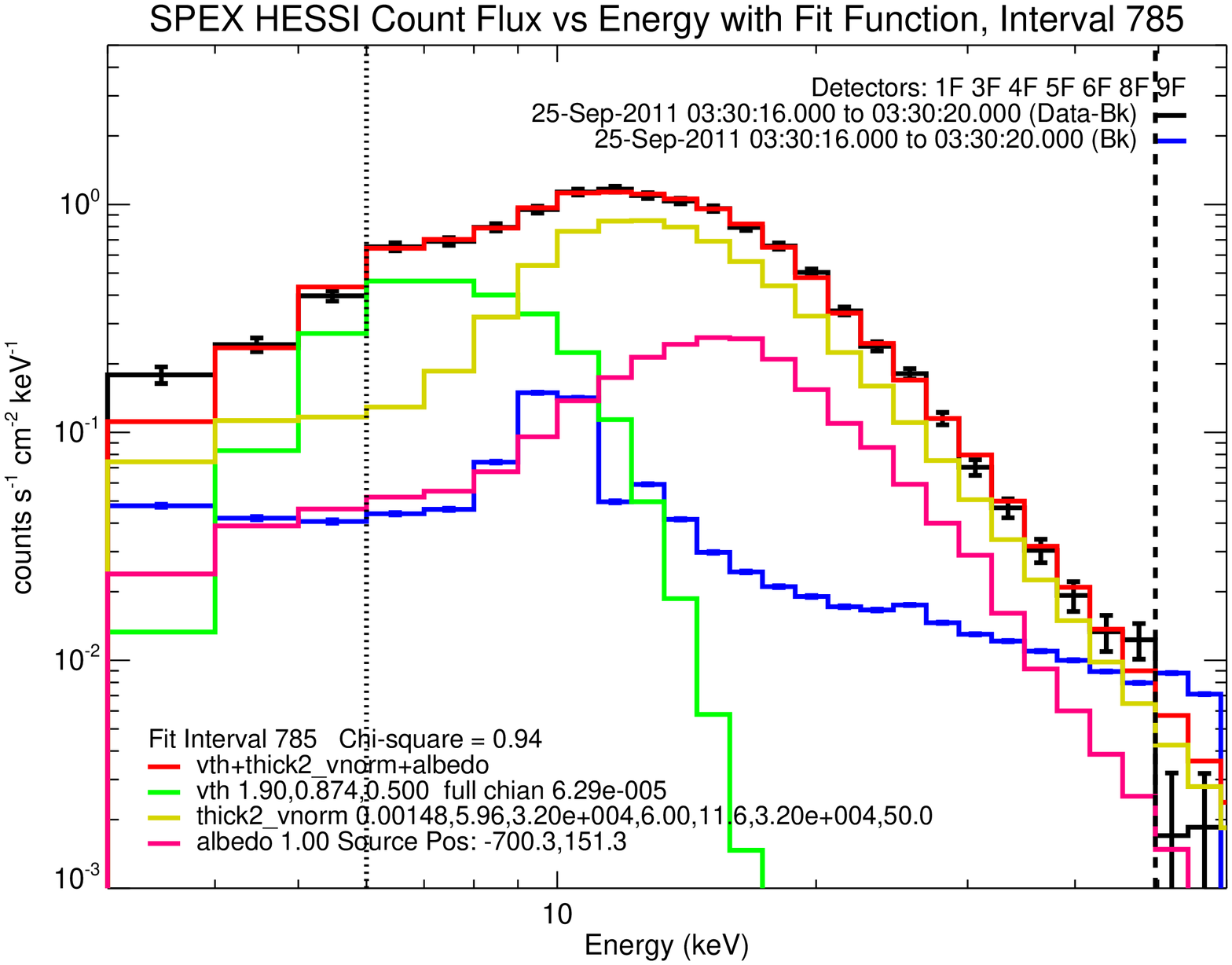}
     \includegraphics*[width=0.5\textwidth, angle = 90, 
   trim = 0 0 25 0]
   		{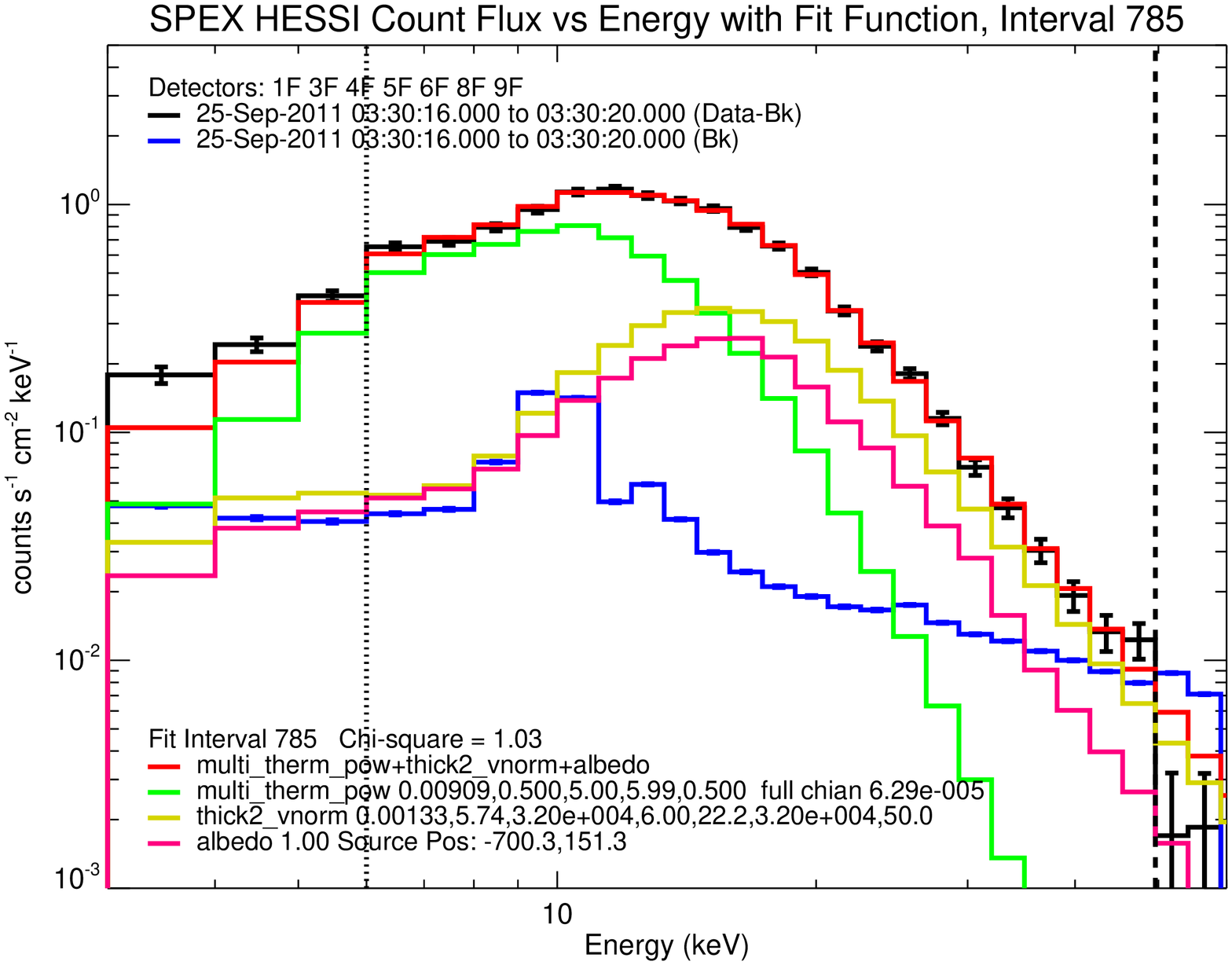}
    \caption{\textbf{Top}: RHESSI measured and fitted count flux spectra summed over the front segments of all detectors except for \#2 and \#7 for a 4~s interval centered on the peak in the 25--50~keV lightcurve at 03:30:18 UT. The measured count fluxes (shown in black with $\pm1\sigma$ statistical uncertainties) are fitted to the sum (shown in red) of a single-temperature thermal function (vth in green), the bremsstrahlung spectrum from electrons with a power-law distribution (thick2\_vnorm in yellow), and the albedo from an assumed isotropic source (in pink).  The blue line is the background spectrum estimated from the pre- and post-flare count rates. \textbf{Bottom}: Same as top plot but with the single-temperature thermal function replaced with a differential emission measure that is a power-law in temperature with an index of 6.0 between temperatures of 0.5 and 5 keV (multi\_therm\_pow in green).  
    }
   \label{Fig-sp-peak}
   \end{figure}

\subsection{Spectroscopy}
    \label{Spectroscopy}
Detailed spectral analysis shows that the source can be detected above background to energies as high as 50 keV during the impulsive peak in the 25--50~keV lightcurve at 03:30:18 UT shown in Figures~\ref{Fig-lc25sep2011} and \ref{Fig-lc-RHESSI-GOESderiv}. The RHESSI count flux spectra for a 4~s interval about that time are shown in Figure \ref{Fig-sp-peak} along with thermal, nonthermal, and albedo spectra that together best fit the measurements. Figure~\ref{Fig-sp-peak}~(top) shows the measured background-subtracted count flux spectrum (black histogram  with $\pm1\sigma$ error bars) between 3 and 50 keV in 0.3 keV bins summed over the front segment of all nine germanium detectors except for \#2 and \#7 since they had poorer energy resolution and sensitivity below $\sim$10~keV. The red histogram is the sum of the following functions giving the best fit (reduced $\chi^2~=~0.94$) to the measured spectrum between 6 and 50 keV: 
    \begin{description}
    \item[vth] Single-temperature thermal spectrum (vth) with an emission measure of $1.9~10^{49}~cm^{-3}$ and a temperature of $0.87$~keV (10~MK). 
    % A multi-temperature function is used in Figure \ref{Fig-sp-peak}(bottom) with a differential emission measure that is a power-law in temperature. 
    The Fe abundance was assumed to be 0.5 times the Chianti chromospheric {value for all the thermal functions used here}.
    
    \item[thick2\_vnorm] A cold thick-target bremsstrahlung X-ray spectrum from a power-law electron distribution with a flux at 50 keV of $1.5\pm0.1~10^{32}$~electrons~$s^{-1}~keV^{-1}$, an index ($\delta$) of $6.0\pm0.1$, and a low energy cutoff ($E_c$) of $\ge12$~keV.
    
    \item[albedo] Predicted albedo spectrum for an isotropic source at the given location on the solar disk following \cite{2006A&A...446.1157K}.
    \end{description}

The best-fit spectrum shows that the contributions of the thermal and nonthermal components are equal at $\sim$10 keV. The value of $\sim$12 keV obtained for $E_c$ {is usually taken to be} an upper limit to its true value since a lower value would give an equally good fit to the data \citep{2013ApJ...769...89I}. {However, if a higher temperature thermal component is also present, this can be a false assumption as indicated below.}

An equally valid fit, shown in Figure ~\ref{Fig-sp-peak}~(bottom), was obtained when the single-temperature function was replaced with a multi-temperature model in which the differential emission measure {(DEM) is a power-law function of temperature of the form
\begin{equation}
DEM(T) = A~(T/2)^{-\alpha}~cm^{-3}~keV^{-1}
\label{eq-DEM}
\end{equation}
where T is the temperature in keV and A is the DEM at T = 2 keV. This function, called multi\_therm\_pow in Figure \ref{Fig-sp-peak}, is restricted to the temperature range between T$_{\min}$ and T$_{\max}$. Fitting this function to the count flux spectrum with A and $\alpha$ as free parameters and T$_{\min}$ and T$_{\max}$ fixed at their default values of 0.5 keV and 5 keV, respectively, gives an acceptable value of $\chi^2$ (=1.03). The best-fit values of A and $\alpha$ are $9.1 \pm0.4~10^{46}~cm^{-3}~keV^{-1}$ and $6.0\pm0.2$, respectively, but the low energy cutoff ($E_c$) is increased from the 12 keV obtained with the single temperature assumption up to 22 keV.}

{We compared this very steep DEM function with the DEM deduced using the method given by \cite{2012A&A...539A.146H} with AIA images taken of this flare at this time in the six channels with temperature coverage above 1 MK (the 94, 131, 171, 193, 211, and 335~\AA~channels). Although there is some uncertainty in the fluxes in some channels because of bleeding from the brightest pixels, particularly in the 171~\AA~ channel, the two functions cross at a DEM of a few times $10^{24}~cm^{-5}~K^{-1}$ (using a source area of $2\times10~arcsec^2=1.2~10^{17}~cm^{2}$) at a temperature of $\sim$1 keV ($\sim$12 MK). AIA does not significantly constrain the DEM at higher temperatures and RHESSI does not constrain it at lower temperatures. Changing $T_{\min}$ from 0.5 to 1~keV has little effect on the goodness of fit to the RHESSI data except for the lowest energy point in the 6-7~keV bin. This point is also directly dependent on the Fe abundance so it is impossible to determine these two parameters independently of one another with attenuators in place. (A better approximation to the expected function would be for the DEM in all multi\_therm functions to be set to a fixed value equal to its value at $T_{\min}$ for all lower temperatures, rather than to zero as is currently the case.)}

{A third possible thermal function - the sum of two single-temperature spectra (vth + vth) with different temperatures - also gave an acceptable fit to the count rate spectrum for this time interval. In this case, the two temperatures were $0.7\pm0.2$ and $1.7\pm0.1$~ keV with corresponding emission measures not well constrained at $4\pm4~10^{49}~cm^{-3}$ and $2.1\pm0.3~10^{47}~cm^{-3}$, respectively.}   

\begin{figure}
    \includegraphics[width=1.0\textwidth, angle = 0, 
   trim = 0 0 0 0]
   		{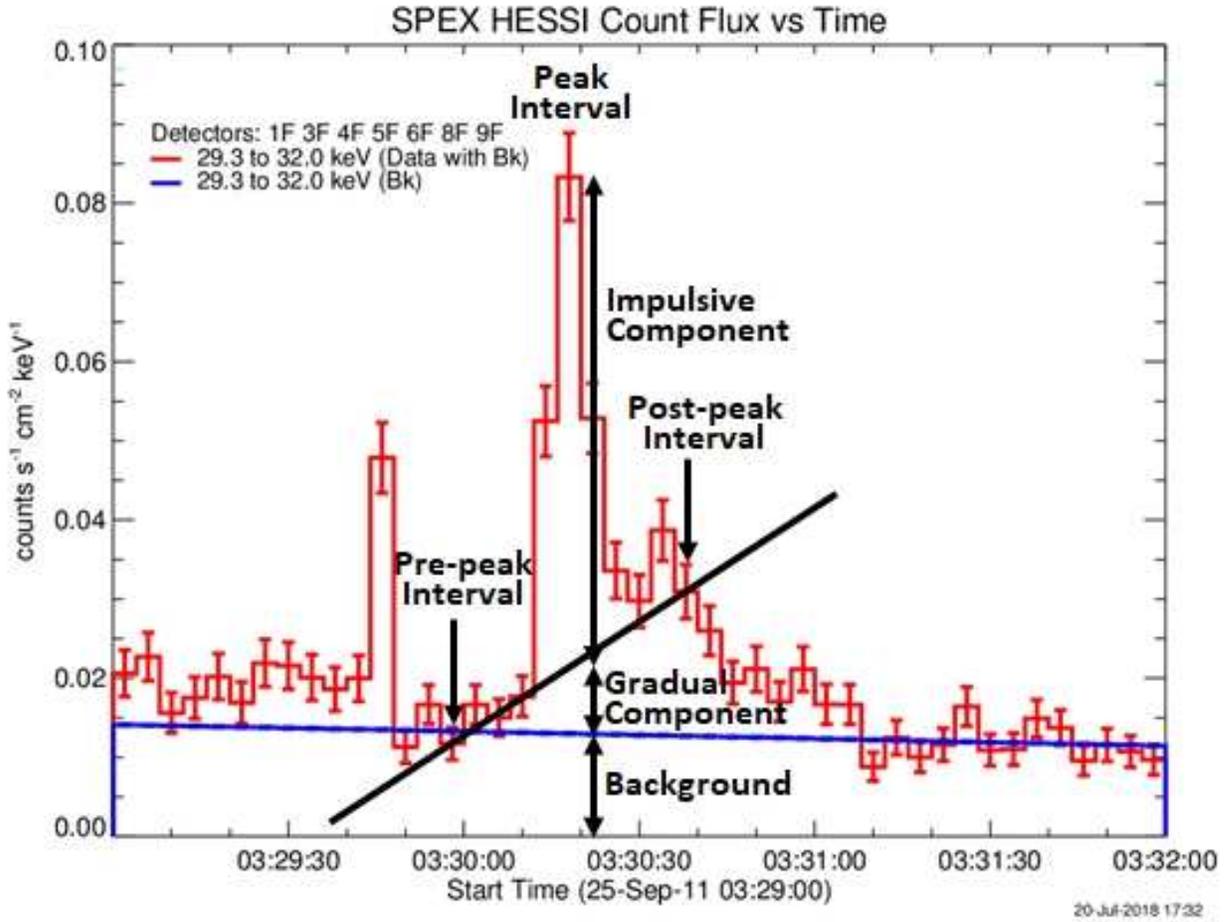}
    \caption{Definition of the impulsive and gradual components of the impulsive peak at 03:30:18~UT. RHESSI 29 - 32 keV light curve showing the count rate flux summed over the front segments of all nine detectors except for \#2 and \#7 with $\pm1\sigma$ statistical error bars. The peak interval and the pre- and post-peak intervals are indicated to show how the background-subtracted impulsive and  gradual components were determined for the plot in  Figure \ref{Fig-peak-ratio} of the ratio of the two for comparison with the ratio of the nonthermal to thermal components of the spectrum in the peak interval. Note that the spurious high point at 03:29:46 UT was caused by the insertion of the thin attenuators at that time.
    }
   \label{Fig-definitions}
   \end{figure}

\begin{figure}
\centering
    \includegraphics*[width=0.5\textwidth, angle = 90, 
   trim = 0 0 20 0]
   		{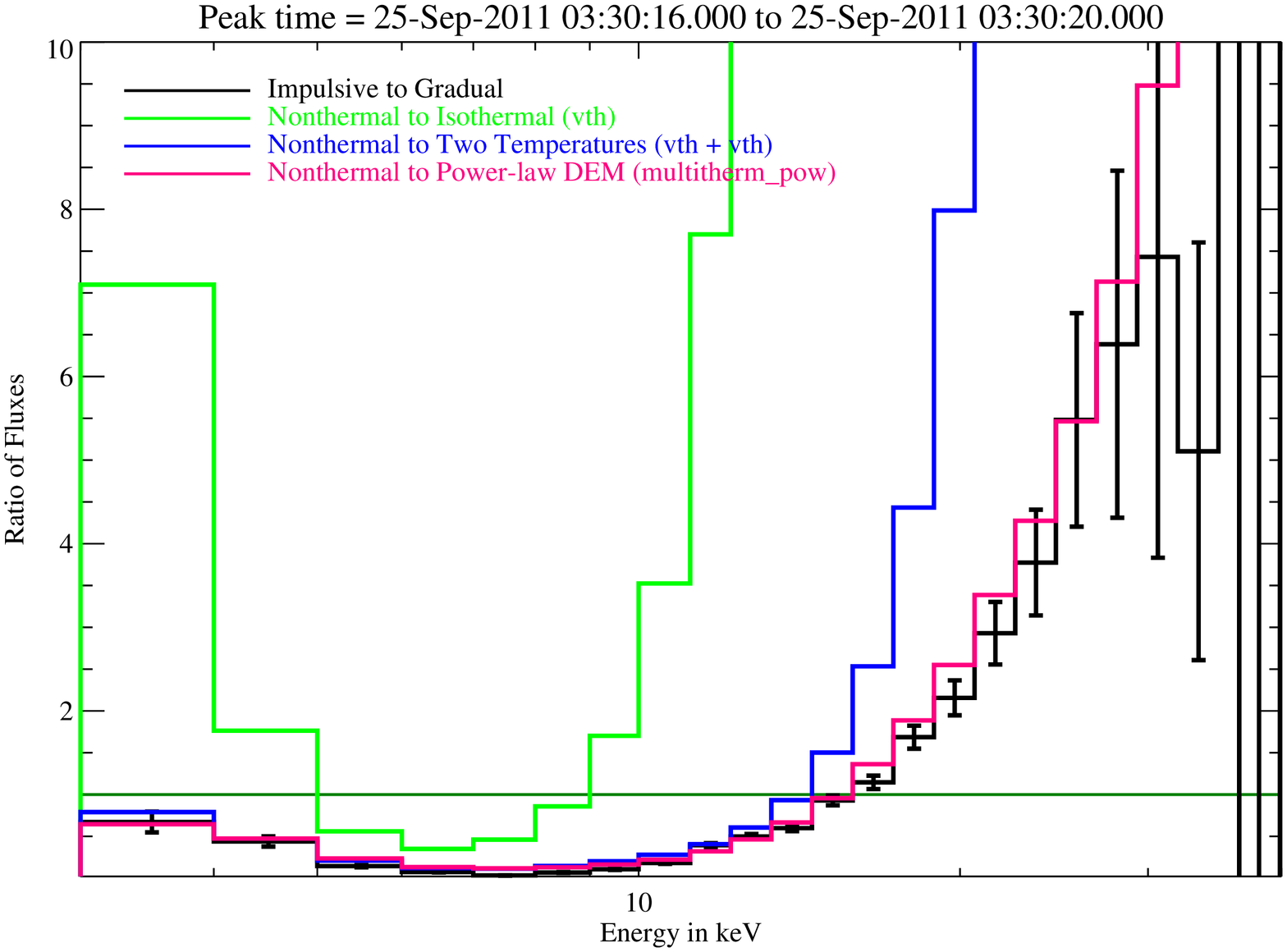}
    \includegraphics*[width=0.5\textwidth, angle = 90, 
   trim = 0 0 20 0]
   		{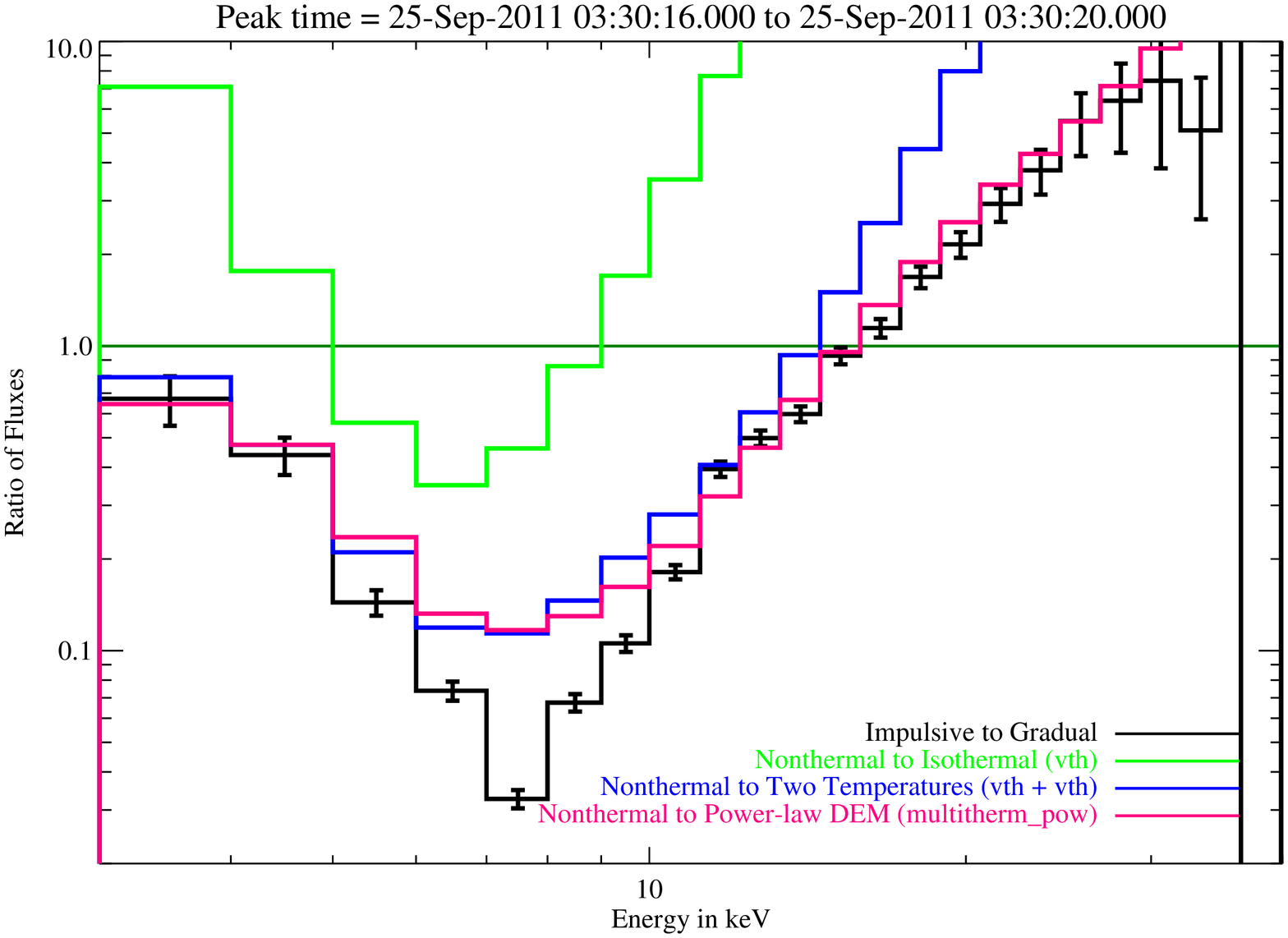} 
    \caption{Comparison of the energy dependence of the ratio of nonthermal to thermal fluxes to the ratio of impulsive to gradual flux variations for the 4~s time interval starting at 03:30:16 UT at the peak in the 25 to 50 keV energy range (see Figure~\ref{Fig-definitions} for definitions of the impulsive and gradual components of the impulsive peak). The two plots are the same except that the Y-axis is linear in the top plot and logarithmic in the bottom plot. The black histogram with statistical error bars shows the ratio of the impulsive-to-gradual emission. The three colored histograms with no error bars show the nonthermal-to-thermal flux ratios with the nonthermal flux determined from the thick2\_vnorm component and the thermal flux from three different forms of the thermal component: a single thermal function (vth) shown in green, a two-temperature function (vth + vth) shown in blue, and a power-law differential emission measure (multi\_therm\_pow) shown in pink. Note that the nonthermal-to-thermal ratio determined from the multi\_therm\_pow fit agrees closest with the impulsive-to-gradual ratio.
    }
   \label{Fig-peak-ratio}
   \end{figure}

Since both a single {and a double temperature function} and a power-law DEM all give acceptable values of $\chi^2$, there is no way to determine which is correct from spectral analysis alone. One way to differentiate between them is to estimate the magnitudes of the thermal and nonthermal components as a function of energy based on the impulsiveness of the light curves. One would expect that the thermal emission would vary more gradually than the nonthermal emission based on the idea that there is steady heating compared to the impulsive particle acceleration. Also, any impulsive heating by the particles would decay more gradually because of the longer cooling time as expected based on the Neupert Effect \citep{1968ApJ...153L..59N}. The impulsive peak at 03:30:18~UT offers the opportunity to track its relative amplitude as a function of energy as compared to the more gradually varying emission seen at lower energies. The definition of the gradual and impulsive components is shown in Figure~\ref{Fig-definitions}. The gradual component is the count flux above the preflare background level and below a linear interpolation between the fluxes in pre- and post-peak 4~s intervals (20 s on either side of the peak). The impulsive component is the flux above the gradual component. In Figure~\ref{Fig-peak-ratio}, the plotted points with $\pm1\sigma$ statistical uncertainties are the ratio of these impulsive and gradual components as a function of energy. This plot shows that the impulsive peak was not evident in the light curves up to $\sim$10 keV (ratio $\ll1$) consistent with the gradual increase expected for the thermal emission. Above 10 keV, the impulsive peak became more dominant over the gradual component up to $\sim$30 keV as expected for the more impulsive nonthermal emission. Above 30 keV, the measured flux becomes closer to the pre-flare background with the resulting larger error bars. 

Also shown in Figure~\ref{Fig-peak-ratio} for comparison with the impulsive-to-gradual ratios are the thermal-to-nonthermal ratios obtained from the spectral fitting of three assumed thermal functions - isothermal (vth), two temperature (vth$+$vth), and the power-law DEM (multi\_therm\_pow). As can be seen, the ratios determined from both the single and double temperature functions deviate from the impulsive/gradual ratios at the higher energies whereas the ratios from power-law DEM function agrees over all the energies up to at least 30 keV.

The ratios of both the thermal/nonthermal and impulsive/gradual ratios all increase with decreasing energy below $\sim$6 keV. This is presumably because the thin attenuators were moved into the light path of each detector just before the peak, thus strongly attenuating the X-ray fluxes at low energies. Most of the counts recorded with energies below 6 keV are then from higher energy X-rays that are photoelectrically absorbed but the K-shell fluorescence photon (about 10 keV) escapes \citep{2002SoPh..210...33S}. Thus, the recorded counts below $\sim$6 keV can be expected to have the same ratio as for X-rays with 10 keV more energy than the measured energy loss in the detector. This is shown by the ratios for the 3--4 keV X-rays being the same as for the 13--14 keV X-rays.

This timing analysis thus supports the conclusion reached from the spectral analysis that the thermal and nonthermal emission contribute equally at $\sim$15 keV and that the low energy cutoff cannot be much below 20 keV. If it were, one would expect the impulsive peak to be evident to lower energies. This is similar to the conclusion reached by \cite{2005ApJ...626.1102S} based on their method combining spatial, spectral, and temporal analysis to determine the cutoff energy but they did not consider the effects of a non-isothermal plasma {contributing significantly to the X-ray spectrum at higher energies}.

\section{\textbf{Discussion}}
    \label{discussion}
    
We have presented detailed analysis of an unusually narrow X-ray source with an FWHM width close to the 2 arcsec resolving power of the finest RHESSI grids. Other small sources imaged with RHESSI have been reported by \cite{2009ApJ...698.2131D} but they were usually in pairs at energies above $\sim$25 keV and were interpreted as footpoint sources with some extent along the corresponding ribbon. There is no such unambiguous interpretation of the RHESSI observations for this event. This narrow source is detected not just in the nonthermal hard X-ray energies above 25 keV but, as shown in Table~\ref{Tab-LoopWidth-Energy}, extends in energy down to at least 6 keV where the counts are dominated by thermal emission, as shown in Figures~\ref{Fig-sp-peak} and  \ref{Fig-peak-ratio}. 

The comparison with the HMI magnetic field image in Figures~\ref{Fig-im_HMI_RHESSI} and \ref{Fig-im_HMIMAG_AIA1700_RHESSI} shows that the RHESSI source lies along the eastern end of the neutral line between two strong and uniform magnetic fields. The AIA 1700~\AA~contours in Figure~\ref{Fig-im_HMIMAG_AIA1700_RHESSI} show what may be interpreted as a long ribbon in the region of positive magnetic polarity coincident with the RHESSI source. The second more compact source is in the region of higher negative magnetic field strength. It is at the same location as a possible weak compact source seen in the RHESSI image made with the MEM NJIT reconstruction algorithm shown in Figure~\ref{Fig-image1}. Perhaps this second compact ribbon is so weak in X-rays because the accelerated electrons are mirrored by the stronger magnetic field resulting in weaker bremsstrahlung emission. AIA images at other wavelengths show a narrow source extending from the two ribbons with an extension in a northerly direction not seen with RHESSI.

{Thus, one interpretation of this event is that very low-lying loops extend from the compact ribbon in the west to different footpoints along the linear ribbon to the east, with further extensions to the north.} The thermal loops seen with RHESSI at energies below $\sim$15 keV happen to overlay the footpoints seen at higher energies and so the same long narrow source is seen at all energies.

{Another possible interpretation is that the X-ray source is of sufficiently high density that the accelerated electrons deposit all of their energy in the coronal part of the loop or loops before reaching the footpoints - the assumption made originally by \cite{Guo2012,Guo2012a,Guo2013,Guo2014}.
To test this model, we estimated the density (n) in the X-ray source from the emission measure (EM) determined from the RHESSI spectral analysis and the source volume (V) from the RHESSI images. The best-fit to the measured spectrum for the single temperature assumption (vth) gives EM~=~$1.9~10^{49}~cm^{-3}$ (Figure~\ref{Fig-sp-peak}). The source volume under the assumption of a single horizontal cylinder with the diameter of the circular cross section equal to the measured FWHM source width of 2 arcsec and the measured length of 10 arcsec is $V=\pi/4\times2^2\times10~arcsec^3=31\times(7.6~10^7)^3~cm^3=1.4~10^{25}~cm^3$.
Using the relation, $EM=n^2~V$ gives $n=1.2~10^{12}~cm^{-3}$}. 

{Calculating the density from the multithermal model (multi\_therm\_pow) is more complicated since it depends on the ill-defined minimum temperature. The integrated emission measure (EM) from $T_{min}$ to $T_{max}$ is given by
\begin{equation}
EM = \frac{A~2^{\alpha}}{(\alpha - 1)} \left[ \frac{1}{T_{min}^{\alpha-1}} - \frac{1}{T_{max}^{\alpha-1}} \right]~cm^{-3}
\label{eq-EM}
\end{equation}
For $T_{min}=0.5~keV$ and $T_{max} = 5~keV$, $EM = 1.2~10^{49}~cm^{-3}$, giving a density of $n = 1.4~10^{12}~cm^{-3}$, similar to the density obtained for the single temperature case. Increasing $T_{min}$ to 1 keV decreases the density to $3~10^{11}~cm^{-3}$.}
    
{With these estimates of the plasma density and source length, we can calculate the energy needed for an electron to reach the footpoints before losing all of its energy to Coulomb collisions. This energy can be calculated from the following relation \citep[Eq.~2.4]{2011SSRv..159..107H} of the evolution of the electron energy (E) with plasma electron column density ($N_e$) :
\begin{equation}
E^2 = E^2_0 - 2KN_e
\label{eq-E}
\end{equation}
where  $E_0$ is the initial electron energy and $K = 3~10^{-18}~keV^2~cm^2$ assuming that the Coulomb logarithm is 23. We assume that all the electrons are injected at the center of the X-ray source and that the density is the same at all locations within the source so that $N_e$ is equal to the source density times half the source length. For a single temperature source, this energy is then 49 keV, while for the multi\_therm\_pow case, the energy is 57 keV. Thus, in either case, most of the electrons producing the nonthermal X-ray emission would be stopped in the loop before reaching the footpoints, thus explaining why no HXR footpoints were detected. In fact, examination of the AIA 1700~\AA~images at the time of the HXR impulsive peak does show bright saturated pixels at the purported location of the western footpoint suggesting that a significant flux of accelerated electrons does reach that footpoint but not enough to be detectable with the RHESSI imaging capability.}

{Another test of the thick-target coronal loop model is the appearance of the flare in the images of the different AIA channels. The 94, 131, and 193~\AA~images should show emission from the hot $\ge$10~MK plasma with more spatial detail than is possible with RHESSI. The 94~\AA~images at the time of the HXR peak show bright areas over the two purported ribbons seen in the 1700~\AA~images. The 131~\AA~images show saturated pixels at the location of the RHESSI source as expected but with an extension to the western bright points seen in the 1700~\AA~images and a weaker extension to the North. The 193~\AA~images show saturated pixels all along the bright ribbons seen in the 1700~\AA~images including at the location of the RHESSI source, again suggesting that some high-energy electrons do reach the footpoints of the loops.
}

\section{Conclusions}
	\label{conclusions}

{The unusually narrow X-ray source imaged with RHESSI during an impulsive spike lasting for $\sim$10~s during the GOES C7.9 flare on 25 September 2011 (SOL2011-09-25T03:32) was only $\sim$2~ arcsec wide and $\sim$10~ arcsec long. Comparison with HMI magnetograms and AIA images at 1700~\AA~shows that the X-ray emission was primarily from a long ribbon in the region of positive polarity with little if any emission from the negative polarity ribbon. It is not clear why the X-ray source is located over this ribbon although it may be that the negative magnetic field was stronger and/or with greater convergence at that location to cause more magnetic mirroring.}  

{The absence of clear double footpoints explains why this event was selected by \cite{Guo2012a,Guo2012,Guo2013} as a coronal hard X-ray flare. Indeed, a thermal plasma source density of $\sim$10$^{12}~cm^{-3}$ estimated from the RHESSI-derived emission measure and source area shows that this could best be interpreted in this way such that electrons accelerated in the corona to energies of less than $\sim$50~keV would be stopped by Coulomb collisions before reaching the footpoints. However, as pointed out by \cite{2018ApJ...867...82D}, the measured increase in source length used by these authors to determine various parameters of the source region is probably incorrect since it does not take into account the presence of weak footpoints at higher energies and the difficulty associated with determining the source length in a loop viewed from almost directly overhead.}

This exceptionally narrow source has afforded the opportunity to test the ability of RHESSI's finest grids to modulate the incident X-ray flux with the expected amplitude. While such modulation was demonstrated before launch, the fact that very few flares have shown any modulation in the two detectors behind the grids with the finest pitch brought into question the continued alignment of the two finest grid pairs after launch.  In retrospect, it is not surprising that most HXR sources are $>$ a few arsec in extent since even a point source near disk center at an altitude near the top of the chromosphere ($\sim$3~Mm or 3~arcsec above the photosphere) will appear as an extended source with a FWHM of many arcsec because of the albedo component \citep{2010A&A...513L...2K}. Separating the direct source from the albedo component has proved to be very difficult despite the detection by \cite{2002SoPh..210..273S} of ``core-halo'' structures with ``core'' sizes of $\le$6 to 14~arcsec and ``halo'' sizes of $\sim$40 arcsec.

Another capability afforded by this highly impulsive event is the evaluation of the relative fractions of thermal and nonthermal emission as a function of energy. This is based on the assumption that the impulsive peak lasting for just three RHESSI 4-s rotation periods was from a power-law nonthermal distribution of electrons and the more gradually varying component was from a thermal distribution.  With these assumptions, it was possible to show that the spectrum at the time of the impulsive peak was not consistent with either a single or a two-temperature thermal function but was consistent with a power-law differential emission measure as a function of temperature. The thermal and nonthermal components were of equal intensity at a photon energy of 15 keV. 

The upper limit on $E_c$ (the low energy cutoff to the electron spectrum) determined for the power-law DEM model was 22 keV compared to 12 keV for the single temperature assumption.  Since the total energy in electrons scales inversely with $E_c^{\delta-2}$, this factor of $\sim$1.8 increase in $E_c$ results in a decrease in the lower limit on the total energy in electrons by an order of magnitude given that the spectral analysis gives $\delta = 5.8\pm0.2$ for any of the three thermal assumptions. A similar result was found by \cite{2015ApJ...813...32D} in analyzing a different impulsive event. This shows that it is always important to consider a multi-thermal model in quoting a lower limit to the total energy in electrons since assuming an isothermal model can result in a order of magnitude error in that estimate. This is pointed out in a recent paper by \cite{2019arXiv190605835A}, who also discuss three other possible models for establishing more accurate values of the low energy cutoff that are beyond the scope of this paper. 

A new method has been suggested by \cite{2019ApJ...871..225K} for determining the total accelerated electron rate and power using a ``physically self-consistent warm-target approach which involves the use of both hard X-ray spectroscopy and imaging data.'' They claim that for a flare observed with RHESSI, they can determine not just an upper limit to the low-energy cutoff but also a lower limit so that a value is determined with a $\pm$7\% uncertainty at the 3$\sigma$ level, an accuracy never previously claimed. It remains to be seen if this estimate of the uncertainty is realistic and if the multithermal nature of the hot plasma found for the flare studied in this paper can be incorporated into their analysis. They use a single temperature fit to the RHESSI spectrum in their Figure 2, and, although they give a reduced chi-squared value of 1.18, they have used unreasonably large systematic uncertainties on the measured count rates and the residuals show large systematic variations at energies below $\sim$20~keV, suggesting that a multithermal model is required. 

%\section{ACKNOWLEDGEMENTS}
%\vspace{5mm}
\bigskip
We thank the anonymous referee for their careful reading of the original manuscript and for suggesting that the coronal thick-target interpretation may well be viable for this event. We thank Richard Schwartz for his contributions to the RHESSI software and for his invaluable help in using it. We thank Meriem Alaoui for help with the analysis, and S{\"a}m Krucker and Karen Muglach for critically reading an earlier version of the manuscript and for their many suggestions for improvements.

\section{Appendix - Comparison of Image Reconstruction Algorithms}
\label{Sec-Appendix}
%\subsection{}
%\label{Sec-ImageAllgorithms}

This event provides an opportunity to compare the performance of the different RHESSI image reconstruction algorithms when there is clear evidence of significant modulation in the count rates of detectors \#1 and \#2. The RHESSI imaging concept is described by \cite{2002SoPh..210...61H} with brief descriptions of the image reconstruction algorithms that were available immediately following the launch of RHESSI in February 2002. More algorithms are now available in RHESSI software\footnote{
https://hesperia.gsfc.nasa.gov/rhessi3/software/imaging-software/image-algorithm-summary/index.html}, particularly using visibilities computed from the modulated count rates in each detector. A visibility is defined as a vector representation of the amplitude and phase of the modulation at a given spacecraft roll angle with most instrumental artifacts removed \citep{2002SoPh..210...61H}.

We have made images of this particular flare using all of the currently available algorithms and evaluated their performance in each case. The following is a discussion of the source dimensions obtained with the different algorithms and the relative quality of the images in terms of the agreement between the measured count rates or visibilities and the predicted values from the reconstructed images. Images were made with each algorithm for the same time interval (03:30:08 to 03:31:12 UT) and energy bin (12--25~keV) using the default settings in each case unless indicated otherwise. The result are summarized in Table~\ref{Tab-LoopWidth}.

% Deviations from default settings included the following:
% \begin{itemize}
% \item \textbf{VIS\_CONJUGATE} was set to sum conjugate values of  the visibilities, i.e. those that differ by $180^\circ$ in roll angle.
% \item \textbf{VIS\_NORMALIZE} was set to correct visibility amplitudes to equalize the count rate in each detector averaged over a full spacecraft rotation. Note that this only applies to the algorithms that use visibilities.  Thus, it was not used for CLEAN or Pixon.
% \item \textbf{MPAT\_COORD} was set to `CART' to use Cartesian coordinates instead of `ANNSEC' to use the annular sectors that were used originally to save on computer memory requirements.
% \end{itemize}

Values of the source intensity and the FWHM width and length obtained with each algorithm are listed in Table~\ref{Tab-LoopWidth}. The intensity was obtained by simply summing the flux over all 0.5 arcsec square pixels in the 129x129 array used for each image.  An indication of the background flux outside the main source was obtained by summing the flux in pixels with $>$10\% of the peak flux and expressing that as the listed fraction of the total intensity. This fraction gives an indication the extent of weak fluxes away from the main source region. The image made  with VIS\_FWFIT is expected to have the largest such fraction since it is based on the assumption that there is no emission outside of the assumed Gaussian source. 

The source width and length were obtained using the IDL procedure HSI\_GET\_EXTENT.PRO. It computes the standard deviations $\sigma$s for a source region rotated through 180$^\circ$ in 1$^\circ$ increments and summed over one dimension.  Before this procedure was called, all pixels with fluxes less than 10\% of the peak flux were zeroed to reduce the effects of small fluxes in the images far from the main source. The minimum and maximum values of the calculated $\sigma$s were converted to FWHM values (FWHM~=~$2\sqrt{2ln2}\times\sigma = 2.35\sigma$) and listed as the source width and length, respectively.

Also listed in Table~\ref{Tab-LoopWidth} for each algorithm are two measures of how good the reconstructed images would predict the observations. These are the C-statistic relating the  measured and predicted modulation count profiles as a function of regularized roll angle (see Figure~\ref{Fig-profiles}) and the reduced $\chi^2$ values derived from the relation between the measured and predicted visibilities (see Figure~\ref{Fig-visibilities}).

Three images were made using the VIS\_FWDFIT image reconstruction algorithm with the assumptions that the source was either a circular or an elliptical Gaussian or a loop structure with a Gaussian distribution along and perpendicular to its length. This algorithm is unique in providing the listed statistical $\pm1\sigma$ uncertainties on the source dimensions based on a Monte Carlo analysis using 20 randomly chosen sets of visibilities distributed about the best-fit visibilities according to their statistical uncertainties. 

\begin{table}
\centering                  
\begin{tabular}{||c|c|c|c|c|c|c||}                                           
				\hline
				\hline   
	Algorithm & Flux & Width & Length & Peak  & Cash & Reduced   \\
	          & (ph. $cm^{-2} s^{-1}$) & (arcsec) & (arcsec) &   Fraction         & Statistic    &   $\chi^2$                \\
	\hline\hline
      VIS FWDFIT (circle)     &    125     &     9.5     &     9.5     &    0.90     &     1.7     &    3.54     \\ \hline
                              &    125$\pm$2 & 11.0$\pm$0.4 & 11.0$\pm$0.4 &          &    1.9     &    2.08     \\ \hline
     VIS FWDFIT (ellipse)     &    125     &     1.9     &    14.0     &    0.90     &     1.6     &    1.77     \\ \hline
                              &    124$\pm$2 & 2.10 $\pm$0.27 & 15.8 $\pm$0.44 &     &     1.6     &    1.46     \\ \hline
        VIS FWDFIT (loop)     &    125     &     2.2     &    14.1     &    0.86     &     1.5     &    1.70     \\ \hline
                              &    125$\pm$1 & 2.02$\pm$0.18 & 15.9$\pm$0.4 &         &    1.6     &    1.41     \\ \hline               
               MEM NJIT       &    150     &     1.6     &    10.4     &    0.46     &     1.2     &    1.25     \\ \hline
                   MEM GE     &    146     &     2.4     &     9.9     &    0.47     &     1.2     &    1.62     \\ \hline
               UV\_Smooth     &    211     &     4.1     &    10.8     &    0.48     &     5.6     &    5.59     \\ \hline
                       EM     &    157     &     2.0     &     9.6     &    0.50     &     2.5     &    1.46     \\ \hline
                  VIS\_CS     &    144     &     5.5     &    12.0     &    0.74     &     1.2     &    2.08     \\ \hline
                  VIS\_WV     &    153     &     5.9     &    12.8     &    0.84     &     1.4     &    2.09     \\ \hline
                    Pixon     &    157     &     2.3     &    11.7     &    0.54     &     2.5     &    1.51     \\ \hline
\hline
\multicolumn{7}{||c||}{CBWF $=$ 1.0} \\ \hline
            Clean disable     &    246     &    11.3     &    17.7     &    0.86     &     3.7     &    2.32     \\ \hline
         Clean full\_resid     &    248     &    11.2     &    17.6     &    0.86     &     3.5     &    2.29     \\ \hline
       Clean scaled\_resid     &    129     &     7.1     &    13.3     &    0.87     &     3.6     &    2.29     \\ \hline
   Clean old\_scaled\_resid     &    144     &     6.2     &    12.2     &    0.89     &     3.8     &    2.64     \\ \hline
           Clean no\_resid     &    130     &     7.1     &    13.3     &    0.87     &     3.5     &    2.29     \\ \hline
         Clean media\_mode     &    128     &     7.1     &    13.3     &    0.87     &     3.7     &    2.32     \\ \hline
\hline
\multicolumn{7}{||c||}{CBWF $=$ 2.0} \\ \hline    
%             Clean disable     &    247.     &     6.2     &    13.7     &    0.55     &     3.6     &    1.61     \\ \hline
%          Clean full\_resid     &    248.     &     6.2     &    13.7     &    0.55     &     3.5     &    1.60     \\ \hline
%       Clean scaled\_resid     &    129.     &     2.9     &    12.4     &    0.71     &     3.6     &    1.60     \\ \hline
%   Clean old\_scaled\_resid     &    142.     &     2.8     &    12.1     &    0.74     &     3.7     &    1.90     \\ \hline
           Clean no\_resid     &    130.     &     2.9     &    12.4     &    0.71     &     3.5     &    1.60     \\ \hline
        %  Clean media\_mode     &    128.     &     2.9     &    12.4     &    0.71     &     3.6     &    1.61     \\ \hline
 	            \hline
% \multicolumn{7}{||c||}{Clean\_beam\_width\_factor $=$ 4.0} \\ \hline    
%           Clean disable     &    247.     &     1.6     &    12.7     &    0.39     &     3.6     &    1.37     \\ \hline
%          Clean full\_resid     &    248.     &     1.6     &    12.7     &    0.39     &     3.5     &    1.37     \\ \hline
%       Clean scaled\_resid     &    129.     &     1.5     &    11.6     &    0.64     &     3.6     &    1.37     \\ \hline
%   Clean old\_scaled\_resid     &    142.     &     1.5     &    11.5     &    0.66     &     3.7     &    1.70     \\ \hline
%           Clean no\_resid     &    130.     &     1.5     &    11.6     &    0.64     &     3.5     &    1.37     \\ \hline
%          Clean media\_mode     &    128.     &     1.5     &    11.6     &    0.64     &     3.6     &    1.37     \\ \hline
%  	            \hline
\multicolumn{7}{||c||}{Point source components only (CBWF$=$ 10.0)} \\ \hline
%             Clean disable     &    251.     &     1.1     &    11.3     &    0.28     &     3.4     &    1.37     \\ \hline
%        Clean full\_resid      &    248.     &     1.1     &   14.4      &    0.25     &     3.5     &    1.35     \\ \hline
%       Clean scaled\_resid     &    130.     &     0.7     &    11.1     &    0.51     &     3.6     &    1.35     \\ \hline
%  Clean old\_scaled\_resid     &    142.     &     0.7     &    11.1     &    0.52     &     3.7     &    1.74     \\ \hline
            Clean no\_resid     &    128.     &     1.9     &    13.3     &    0.76     &     3.5     &    1.39     \\ \hline
         \hline

\end{tabular}   
\caption{Values of the total flux in the source, the FWHM source width and length obtained using the IDL procedure HSI\_GET\_EXTENT with the images made with the different reconstruction methods for the 12 - 25 keV energy range and the time interval from 03:30:08 to 03:31:12 UT chosen to include the spike in the 25 - 50 keV light curve shown in Figure \ref{Fig-lc25sep2011}. The additional values given for the VIS\_FWDFIT images with $\pm1\sigma$ uncertainties were obtained from the Monte Carlo analysis incorporated in that procedure. The image obtained using the MEM NJIT image reconstruction algorithm is shown in Figure~\ref{Fig-image1}. The peak fraction is the fraction of the counts remaining after all pixels with fluxes less that 10\% of the peak flux were zeroed. See text for descriptions of the two parameters, the C-statistic and the reduced $\chi^2$, that quantify how well the count rates predicted from the reconstructed images match the measured count rates and the visibilities, respectively. The results using the Clean reconstruction method with different control parameters shown in the bottom half of the table are described in the text. 
    }
\label{Tab-LoopWidth}
\end{table}

\begin{itemize}

\item 
\textbf{VIS\_FWDFIT}\footnote{https://hesperia.gsfc.nasa.gov/rhessi3/software/imaging-software/vis-fwdfit/index.html} is {a visibility-based version of the forward-fitting method described in \cite{2002SoPh..210...61H}.} It starts by assuming that there are {a number of sources (usually $\le$3)} in the chosen field of view that can be any combination of circular, elliptical, or curved loop-shaped Gaussians. {Some or all of the parameters} describing each source (the flux, location, size, and orientation of an elliptical or loop-shaped Gaussian) are adjusted until the model-predicted visibilities agree best with the measured visibilities in a $\chi^2$ sense.  This approach offers fast determinations of these source parameters. However, as with any forward-fitting method, it is critical that the assumptions concerning the number of sources and their shapes match reality.

Both an elliptical source and the loop model give excellent fits to all visibility amplitudes and modulation profiles. The values of the source length and width with uncertainties are listed in Table~\ref{Tab-LoopWidth}. C-statistic and $\chi^2$ values are both in the acceptable range except for the value of 3.54 obtained with the circular Gaussian source assumption. At 2.02$\pm$0.25 arcsec, this algorithm with the loop model gives close to the lowest value for the source width of any of the reconstruction algorithms.  The peak fraction is also the highest at 0.86--0.90 because no "noise" is allowed outside the range of the assumed sources.

\item
\textbf{MEM\_NJIT}\footnote{https://hesperia.gsfc.nasa.gov/rhessi3/software/imaging-software/mem-njit/index.html} \citep{2007SoPh..240..241S} is a Maximum Entropy Method (MEM) algorithm based on visibilities that was originally developed at New Jersey Institute of Technology (NJIT). It provides the best fit to the visibility amplitudes especially for the subcollimators with the finest pitch as indicated in Figure \ref{Fig-visibilities}. This is borne out by the relatively small value of $\chi^2$ of 1.25.  The similarly small value of the C-statistic (1.19) shows the good agreement between the predicted and measured count rates shown in Figure~\ref{Fig-profiles}.  As a result of this close agreement with the visibilities for the finest subcollimators, MEM\_NJIT gives the smallest value for the source width of any algorithm. The relatively small value of the peak fraction of 0.46, on the other hand indicates that there is significant low-level emission from locations away from the main source, a contention that is borne out by images at lower energies that show significant emission to the north-east of the main source in agreement with the bright extension at this location seen in AIA images at this time. 

\item
\textbf{MEM\_GE}\footnote{https://hesperia.gsfc.nasa.gov/rhessi3/software/imaging-software/mem-ge/index.html} uses a Maximum Entropy Method with visibilities that has recently been added to the RHESSI software. It is similar to MEM\_NJIT but more robust against the problem of breakup into multiple compact sources that can occur with MEM\_NJIT. As a result, for this particular time interval, it shows less detailed structure than MEM\_NJIT with a factor of $\sim$2 smaller visibility amplitudes for the finer subcollimators leading to a finest source dimension of up to a factor of 2 larger than that obtained with other algorithms. The peak fraction and the C-statistic are comparable to those obtained with MEM\_NJIT but the $\chi^2$ value is higher, commensurate with the poorer reproduction of the visibility amplitudes for the finer subcollimators.

\item
\textbf{UV\_smooth}\footnote{https://hesperia.gsfc.nasa.gov/rhessi3/software/imaging-software/uv-smooth/index.html}$^,$\footnote{https://hesperia.gsfc.nasa.gov/rhessidatacenter/imaging/uv\_smooth-documentation.pdf} starts with the measured visibilities for the selected detectors. It interpolates these visibilities in the uv plane and reconstructs the image using a fast Fourier transform inversion. Ringing effects are reduced by imposing a positivity constraint \citep{2009ApJ...703.2004M}.

This algorithm gave poor fits to visibility amplitudes for the coarser grids \#6 to 9 and poor fits to the average fluxes in each detector. This results in the large value of the source width that is a factor of two larger than the width obtained by other algorithms. The C-statistic and $\chi^2$ values are also very high.

\item
\textbf{EM}\footnote{https://hesperia.gsfc.nasa.gov/rhessi3/software/imaging-software/em/index.html}, the Expectation Maximization algorithm, is based on the Lucy-Richardson Maximum Likelihood method for solving inverse problems \citep{2013A&A...555A..61B}. It gives one of the smallest source widths but the C-statistic is quite large at 2.46.  The $\chi^2$ value is in the acceptable range. 

\item 
\textbf{VIS\_CS}\footnote{https://hesperia.gsfc.nasa.gov/rhessi3/software/imaging-software/vis-cs/index.html} uses a custom Gaussian basis on the assumption that sources can be represented as linear combinations of a number of Gaussian distributions \citep{2017ApJ...849...10F}.
It produced visibility amplitudes in Detectors \#1 or \#2 that were only $\sim$10\% and 30\% of the measured amplitudes, respectively, even with the ``sparseness'' parameter set to zero. Thus, the source width of 5.5 arcsec is significantly larger that that found with other algorithms. The C-statistic is one of the lowest values obtained since this method deliberately minimizes this quantity. This is acceptable providing that there are no sources out of the field of view or that an extended source does not exist with no modulation by the coarsest subcollimators used in reconstructing the image.

\item
\textbf{VIS\_WV}\footnote{https://hesperia.gsfc.nasa.gov/rhessi3/software/imaging-software/vis-wv/index.html} is a finite Isotropic waVElet transform Compressed Sensing image reconstruction algorithm based on visibilities \citep{2018A&A...615A..59D}.
This algorithm gave poor fits to the visibility amplitudes for subcollimators \#1, 2 \& 3. As a result it gives the large value of 5.9 arcsec for the source width, at least a factor of 2 larger than that obtained by other algorithms. The C-statistic is in the acceptable range but $\chi^2$ is high at 2.09.

\item
\textbf{Pixon}\footnote{https://hesperia.gsfc.nasa.gov/rhessi3/software/imaging-software/pixon/index.html} is an adaptation of the program used to analyze data from Yohkoh/HXT \citep{1996ApJ...466..585M}. It aims to produce the simplest model for the image that is consistent with the data. No normalization of the detector sensitivities based on the measured average fluxes was used.  This results in a larger value for $\chi^2$ and the C-statistic than would be obtained if this normalization were used but does not significantly change the dimensions of the source in the reconstructed image. Pixon gives a source width of 2.3 arcsec, close to the smallest value obtained by any algorithm. The peak fraction is surprisingly low, presumably due to significant weak emission away from the main source. This could indicate that the assumption of a single elliptical or loop structure assumed for VIS\_FWDFIT is not adequate and may explain the larger values of $\chi^2$ for that algorithm. 

\item
\textbf{CLEAN}\footnote{https://hesperia.gsfc.nasa.gov/rhessi3/software/imaging-software/clean/index.html} is an iterative algorithm based on the assumption that the image can be well represented by a superposition of multiple point sources \citep{2002SoPh..210...61H}. We have explored various ways of estimating the source parameters using the basic CLEAN algorithm and give the values obtained in Table~\ref{Tab-LoopWidth}. The C-statistic and $\chi^2$ values were computed in the same way as for all the other reconstruction techniques.  For each case, we used the the standard cleaning process  in which up to 100 point sources are obtained from the back-projection image. The resulting image is convolved with a Gaussian with a FWHM representative of the combined resolution of subcollimators used in making the image.  In addition, a fraction of the residual image remaining of the dirty map after all the point sources and their side lobes have been subtracted can be added back into the final reconstructed image as an indication of the uncertainties involved and the possibility of a more extended source or sources in the image that cannot be well represented by the 100 CLEAN point-source components.

We found that the source width obtained with CLEAN using the standard default settings (listed as the ``Clean disable'' algorithm in Table~\ref{Tab-LoopWidth} was significantly larger than that obtained by other methods - 11.3 arcsec vs. 1.6 arcsec obtained with MEM\_NJIT. Part of the problem is that the FWHM of the convolving Gaussian is too large. Better agreement with other algorithms can be obtained by dividing the width of the convolving Gaussian by a factor of two using the so-called ``clean beam width factor'' (CBWF) of 2.0. The results listed in Table~\ref{Tab-LoopWidth} show that using CBWF = 2.0 gives significantly smaller source widths than using CBFW = 1.0, as expected, but still factors of 2 or 3 larger than the widths given by other algorithms.  The values of $\chi^2$ are smaller reflecting the better agreement with the visibilities of detectors 1 and 2 but the flux and C-statistic values remain unchanged.

Following \cite{2009ApJ...698.2131D}, we used an alternative method of determining source dimensions by considering the location and intensity of the point sources alone. This is most easily achieved by using CBWF = 10. The resulting source dimensions are listed in the last line of Table~\ref{Tab-LoopWidth}. This method gives a significantly smaller value of the source width (1.9 arcsec vs. 11.3 arcsec using the default settings) that is similar to the values of 2.02$\pm$0.18 and 1.6 arcsec obtained with VIS~FWDFIT and MEM~NJIT, respectively. Much improved values of $\chi^2$ were also obtained (1.39 vs. 2.34). The source flux (307~$photons~cm^{-2}~s^{-1}$) and C-statistic (3.6) were the same since the same regression coefficient was used to normalize the expected and measured counts. 

Several techniques (see below for details of each one) have been devised and are available in the RHESSI imaging software to ensure that the final reconstructed image predicts detector count rates that agree with the measured rates as closely as possible. The different methods are listed for handling the residuals remaining in the back-projection ``dirty'' map after the 100 point-source components and their side lobes are removed. The simplest approach, the so-called ``media\_mode,'' is to not add the residuals to the final image at all.

 These are handled with the clean\_regress\_combine control parameter that has five choices (prior to April 2018, it was an on/off switch).  The options now are as follows:

\begin{enumerate}

\item disable - no regression, residual map added to component map or not, according to media mode setting (this is the default)

\item full\_resid  - Regression of the component map expected counts against the observed. Scale the component map by the regression coefficient and add the unscaled residual map

\item scaled\_resid - First do the steps for `full\_resid' but then take the counts minus the scaled expected counts and regress that against the expected counts from the residual map. Form the new map from the scaled component map added to the newly scaled residual map

\item old\_scaled\_resid - Only method available prior to April 2018. Regression on the expected counts from the component and residual maps.

\item no\_resid - Perform the regression of `full\_resid', scale the component map by the regression coefficient and do not add any of the residual map. This has the same effect as the so-called `media mode' with the component map scaled to match the measured count rates.

\end{enumerate}

In the preferred methods (full\_resid or no\_resid), the predicted counts from the component map are regressed against the observed counts and then scaled by the regression coefficient. In the full\_resid option, the unscaled residual map is added to the new component map but this gives a higher source flux than the other algorithms if integrated over the whole image. Better agreement is obtained by not including the residuals in estimating the source flux as in the ``no\_resid'' or ``media\_mode'' options. 

In all cases except when no residuals are included in the image, negative values appear in pixels away from the main source. In those cases, the listed source flux and peak fraction values were calculated using only the positive values.

\end{itemize}

In conclusion, it appears that the ``best'' values for the parameters for the 12--25 keV source as determined using HSI\_GET\_EXTENT are a source width between 1.6 arcsec (MEM\_NJIT) and 2.3 arcsec (MEM\_GE) and a length between 9.6 (EM) and 14.1 arcsec (VIS\_FWDFIT loop). This is in agreement with the values and uncertainties for the loop width obtained directly with VIS\_FWDFIT  (2.02$\pm$0.25 arcsec) but not with the loop length (15.91 $\pm$0.48).  This is presumably because of the different assumptions used in the two cases. 

Based on these results, we can provide advice on which reconstruction algorithms are most able to accurately characterize this particular source and presumably most images that show fine structure with significant visibility amplitudes in RHESSI's finest subcollimators. VIS\_FWDFIT gives accurate source parameters with uncertainties provided that the assumptions concerning the number and shapes of the sources are correct and complete.  Other algorithms such as MEM\_NJIT, CLEAN, and Pixon can be used to check on those assumptions. It would seem that UV\_SMOOTH, VIS\_CS, and VIS\_WV are not optimized for images with such fine structure. CLEAN is a special case in that it can give the finest source dimensions comparable to those obtained with VIS\_FWDFIT and MEM~NJIT but only if the point-source components are used without convolving them with a Gaussian or adding the residuals left over from the back-projection dirty map.

\bibliography{NarrowFlare,NarrowFlare1}

\allauthors

\end{document}